\documentclass[11pt,a4paper]{article}

\usepackage[T1]{fontenc}
\usepackage[margin=2.5cm]{geometry}

\usepackage{graphicx}
\usepackage{float}
\usepackage{amsmath}
\usepackage{amssymb}
\usepackage{amsthm}
\usepackage{bbm}
\usepackage{algorithm}
\usepackage{algpseudocode}
\makeatletter
\renewcommand\floatc@ruled[2]{{\@fs@cfont\boldmath #2}\par}
\makeatother
\usepackage{booktabs}
\usepackage{cite}
\usepackage{xcolor}
\usepackage[colorlinks=true,linkcolor=blue!60!black,citecolor=green!40!black,urlcolor=blue!60!black]{hyperref}
\hypersetup{pdftitle={Logarithmic Bounds for the Time-Optimal k-Server Problem on the Line with Cumulative Requests},pdfauthor={Hannah Schwede}}
\theoremstyle{plain}
\newtheorem{theorem}{Theorem}
\newtheorem{lemma}[theorem]{Lemma}

\theoremstyle{definition}
\newtheorem{definition}{Definition}

\newcommand{\doi}[1]{\url{https://doi.org/#1}}
\newcommand{\keywords}[1]{\par\medskip\noindent\textbf{Keywords:} #1}

\newcommand{\da}{\mathrel{:=}}
\newcommand{\cost}{\mathrm{cost}}
\newcommand{\nat}{\ensuremath{\mathbb{N}}}
\newcommand{\natp}{\ensuremath{\mathbb{N}^{+}}}
\newcommand{\algA}{\ensuremath{\textup{\textsc{Alg}}}}
\newcommand{\Opt}{\textup{\textsc{Opt}}}
\newcommand{\algone}{\textup{\textsc{Alg}}_1}
\newcommand{\algtwo}{\textup{\textsc{Alg}}_2}
\newenvironment{proofidea}{\par\addvspace{\medskipamount}\noindent\textit{Proof Idea.}\ \ignorespaces}{\par\addvspace{\medskipamount}}

\begin{document}

\title{Logarithmic Bounds for the Time-Optimal \texorpdfstring{$k$}{k}-Server Problem on the Line with Cumulative Requests}
\author{Hannah Schwede\thanks{Department of Computer Science, ETH Zurich, Switzerland.}}
\date{}
\maketitle

\begin{abstract}
In the time model of the $k$-server problem, all servers move simultaneously and the cost of a request is the maximum distance traveled by any server. Even on the line, no deterministic online algorithm is better than $(k+1)$-competitive. We study a variant on the infinite discrete line in which requests are cumulative: a point requested $t$ times must hold $t$ servers. All $k$ servers start at a common origin, there are exactly $k$ requests, and every server serves exactly one request and stays at its point afterwards, for example to carry out a service there. Our model may also help to understand the general time-optimal $k$-server problem.

We show that, in contrast to the general problem, the best deterministic and randomized strict competitive ratios are $\Theta(\log k)$, independently of the largest distance $\alpha$ of a request from the origin. Our lower bounds hold even if $\alpha$ is known, while our algorithm, which keeps the free servers evenly distributed over the current request range, does not need to know~$\alpha$. For every fixed $\alpha$, the best deterministic competitive ratio is $c_\alpha \ln k + O(1)$ with an explicit constant $c_\alpha \in (2, 2+1/\alpha)$. For $k \leq 2\alpha+1$, the leading coefficients of our deterministic bounds differ by a factor of~$2$. A tight bound of $1+H(k-1)$ on the continuous cycle, for instances with a request opposite to the origin, matches the limit of our lower bound on the line as $\alpha \to \infty$ and suggests that this gap stems from the two ends of the line.

\end{abstract}

\section{Introduction}\label{sec:intro}

In the $k$-server problem, a request is served as soon as a server reaches the requested point. For many natural services, however, the server has to stay at the requested point to carry out the service, so it cannot be used for later requests. We study this setting on the line. All $k$ servers start at a common origin, and $k$ requests arrive one by one. A point that is requested $t$ times needs $t$ servers. After each request, an online algorithm has to decide how to move the servers that are still free, without knowing the positions of future requests. All servers move simultaneously, a step takes as long as the largest distance traveled by any server, and the goal is to minimize the total time of all steps. This time model fits settings in which the servers are independent units that move in parallel. The total time divided by $k$ also bounds the average waiting time of a request.

Our model is related to online matching. There, every server also serves exactly one request, but the servers stay at their initial positions until they are assigned to a request. In our model, by contrast, the free servers can move after every request, so an online algorithm can prepare them for future requests. Our algorithm uses this and keeps the free servers evenly distributed over the current request range, so that every possible next request is close to some free server. 

Our model also describes a single period of a server problem with clocks, in which serving servers have to stay at their points until the next clock signal and all servers may move again afterwards.

Finally, our model may help to close the gap for the general time-optimal $k$-server problem, whose best deterministic competitive ratio on the line is only known to lie between $k+1$ and $\frac{k(k+1)}{2}$ (see the related work below). One way to approach the general problem is to split the request sequence into phases with at most $k$ distinct requested points and to analyze each phase separately; it seems hard to understand the general problem without understanding a single phase. Within a phase, it does not seem to help an online algorithm to move a server away from a point that was already requested, since the adversary can request this point again without starting a new phase. Moreover, on the real line, the adversary can request distinct points that are arbitrarily close to each other. If the algorithm keeps a server at each of these points, this resembles requesting one point several times in our model. Hence, a phase of the general problem seems to behave like a variant of our model in which a serving server may leave its point whenever another server takes its place. A natural next step is to study this variant. Our algorithm remains valid there, so our upper bounds still hold (Section~\ref{sec:cycle}). Further steps are to let the servers start in an arbitrary configuration and then to analyze the whole process as a sequence of phases. Even if this does not solve the general problem, it may lead to good online algorithms for request sequences that consist of only a few phases. In this paper, we take the first step and determine the competitive ratio of our model up to constant factors.

\subsection{Our Contributions}
We call this problem the time-optimal $k$-server problem with cumulative requests and study it on the infinite discrete line; Section~\ref{sec:prelim} gives the formal model. Our main result is the following theorem, which we prove at the end of Section~\ref{sec:rand_lb}. To state it, we call an instance an \emph{$\alpha$-instance} if $\alpha$ is the largest distance of a request from the origin; its optimal offline cost is exactly $\alpha$ (Lemma~\ref{lemma:opt_cost}). Moreover, let $H(n)$ denote the $n$-th harmonic number and $\mu_\alpha \da 1/\ln(1+\frac{1}{2\alpha})$.

\begin{theorem}[Main Result]
\label{theo:main}
Let $\alpha, k \in \natp$ with $k \geq 2$, and consider $\alpha$-instances.
\begin{enumerate}
\item[(i)] Every deterministic online algorithm has a competitive ratio of at least $H(k-1)-\frac{1}{\alpha}$, and there is a deterministic online algorithm that does not know $\alpha$ in advance and is $\bigl(3H(k-1)+8\bigr)$-competitive. Every randomized online algorithm has a competitive ratio of at least $\frac14 H(k-1)$. In particular, the best deterministic and the best randomized competitive ratio are both $\Theta(\log k)$, independently of $\alpha$.
\item[(ii)] For every fixed $\alpha$, the best deterministic competitive ratio is $\frac{\mu_\alpha}{\alpha}\ln k + O(1)$ as $k \to \infty$, where the $O(1)$ term depends only on $\alpha$ and $2 < \frac{\mu_\alpha}{\alpha} < 2 + \frac{1}{\alpha}$. The upper bound is attained by the same algorithm as in~(i).
\end{enumerate}
\end{theorem}

Table~\ref{tab:results} summarizes our bounds. For $k \leq 2\alpha+1$, the leading coefficients of our deterministic bounds differ by a factor of~$2$. In Section~\ref{sec:cycle}, we give evidence that this gap stems from the two ends of the line: on the continuous cycle, we prove a tight bound of $1+H(k-1)$, the limit of our lower bound on the line as $\alpha \to \infty$.

\begin{table}[H]
\caption{Bounds on the competitive ratio for $k \geq 2$. Left: non-asymptotic bounds. Right: $\alpha$ fixed and $k \to \infty$, where the $O(1)$ terms depend only on $\alpha$. The randomized upper bounds are attained by our deterministic algorithm.}
\label{tab:results}
\vspace{8pt}
\centering
\renewcommand{\arraystretch}{1.3}
\begin{tabular}{@{}lll@{}}
\toprule
 & Case $k \leq 2\alpha+1$ & Case $k > 2\alpha+1$, fixed $\alpha$ \\
\midrule
Deterministic lower bound & $\frac{2\alpha+1}{2\alpha}H(k-1)$ & $\frac{\mu_\alpha}{\alpha}\ln k - O(1)$ \\
Deterministic upper bound & $\frac{2\alpha+1}{\alpha}\bigl(H(k-1)+2\bigr)$ & $\frac{\mu_\alpha}{\alpha}\ln k + O(1)$ \\
\addlinespace
Randomized lower bound & $\frac{2\alpha+1}{4\alpha}H(k-1)$ & $\frac{2\alpha-1}{4\alpha}\ln k - O(1)$ \\
Randomized upper bound & $\frac{2\alpha+1}{\alpha}\bigl(H(k-1)+2\bigr)$ & $\frac{\mu_\alpha}{\alpha}\ln k + O(1)$ \\
\bottomrule
\end{tabular}
\end{table}

\subsection{Our Techniques}
All our lower bounds start with a request at distance $\alpha$ from the origin, and all further requests lie in $M_\alpha$, the $2\alpha+1$ points within distance $\alpha$ of the origin. We distinguish two phases. In Phase~1, there are at least as many free servers as points in $M_\alpha$. The adversary repeatedly requests a point with the fewest free servers, which forces a movement of cost at least $1$ after few requests. Between two forced movements, the number of free servers is only multiplied by roughly $2\alpha/(2\alpha+1)$, so about $\mu_\alpha \ln k$ movements are forced in Phase~1. In Phase~2, there are fewer free servers than points in $M_\alpha$. With $f$ free servers, some point of $M_\alpha$ is at distance at least $(2\alpha+1)/(2f)-1/2$ from the closest free server. Requesting this point in every step gives a harmonic sum. Our algorithm keeps the free servers evenly distributed over the current request range. In Phase~1, it only redistributes every few steps, each time within time~$1$, and the number of redistributions matches the number of forced movements. In Phase~2, it redistributes after every request at a cost of about $(2\alpha+1)/f$, which again gives a harmonic sum, but loses a factor of~$2$ compared to the lower bound for $k \leq 2\alpha+1$. If $\alpha$ is unknown, our algorithm enlarges its range whenever a request lies outside of it, and all these range increases together cost at most $2\alpha+1$. Our randomized lower bounds follow the same structure, with requests chosen uniformly at random.

\subsection{Related Work}
The $k$-server problem, introduced by Manasse et al.~\cite{manasse_competitive_1988}, has mostly been studied in the distance model, where the cost is the total distance traveled by all servers. They proved a lower bound of $k$ for every metric space with at least $k+1$ points and conjectured it to be tight. The work function algorithm~\cite{chrobak_server_1992,borodin_optimal_1992} is $(2k-1)$-competitive~\cite{koutsoupias_k-server_1995}, the conjecture holds on the line~\cite{chrobak_new_1991} and on trees~\cite{chrobak_optimal_1991}, and Coester et al.~\cite{coester_k-server_2026} recently claimed a proof in general in a preprint.

For randomized algorithms, the best competitive ratio is $\Theta(\log k)$ on uniform metrics~\cite{fiat_competitive_1991}, $\Omega(\log k)$ on every metric space with more than $k$ points~\cite{bubeck_randomized_2023}, improving earlier bounds~\cite{blum_decomposition_1992,bartal_ramsey-type_2006,bartal_metric_2005}, and $\Omega(\log^2 k)$ on some metric spaces~\cite{bubeck_randomized_2023}, while upper bounds polylogarithmic in $k$ and the number of points are known~\cite{bansal_polylogarithmic-competitive_2011,bubeck_k-server_2018}.

The time model has received far less attention. It was introduced by Fiat et al.~\cite{fiat_competitive_1994} as the min-time server problem. Koutsoupias and Taylor~\cite{koutsoupias_cnn_2004} observed that a $c$-competitive algorithm in the distance model is $ck$-competitive in the time model and gave an upper bound of $\frac{k(k+1)}{2}$ on trees, which is optimal for $k=2$ by their lower bound of $3$ on the line. Frei et al.~\cite{frei_time-optimal_2025} proved lower bounds of $2k-1$ for deterministic and $k+H(k)-1$ for randomized algorithms on specific metric spaces, a lower bound of $k+1$ on all Euclidean spaces including the integer line, and a tight bound of $k$ on uniform metrics. Hence, on the line, the best deterministic competitive ratio of the general problem lies between $k+1$ and $\frac{k(k+1)}{2}$.

Our model is also related to online metric matching~\cite{kalyanasundaram_online_1993,khuller_-line_1994}, where $k$ servers are located at fixed points and every request has to be matched immediately and irrevocably to a distinct server, so that, as in our model, every server serves exactly one request. For the total matching distance, the best deterministic competitive ratio on general metrics is $2k-1$~\cite{kalyanasundaram_online_1993,khuller_-line_1994}; on the line, a randomized~\cite{czumaj_online_2012} and a deterministic~\cite{raghvendra_optimal_2018} $O(\log k)$-competitive algorithm are known, while every randomized algorithm has a competitive ratio of $\Omega(\sqrt{\log k})$~\cite{peserico_matching_2023}. For the maximum matching distance, known as online bottleneck matching, \textsc{Permutation} is $(2k-1)$-competitive~\cite{kalyanasundaram_online_1993,anthony_online_2014}, Idury and Sch\"affer showed a lower bound of $\Omega(k)$ even on the line (see~\cite{anthony_online_2014}), and the bound is exactly $k+1$ for equally spaced servers on the line~\cite{xiao_online_2023}. As noted above, the possibility to move the free servers is essential for our bounds, since serving every request directly from the origin can cost up to $k\alpha$. Moreover, our cost is the total time of the simultaneous movements rather than the sum or the maximum of the matching distances, so results do not transfer directly between the two models.

\section{Preliminaries}\label{sec:prelim}
We use the standard framework for online minimization problems and competitive analysis~\cite{komm_introduction_2016}. Let $\mathcal{I}$ be a class of instances. A deterministic online algorithm $\algA$ is \emph{strictly $c$-competitive} on $\mathcal{I}$, for $c \geq 1$, if $\cost(I, \algA) \le c \cdot \cost(I, \Opt)$ for every $I \in \mathcal{I}$, where $\Opt$ denotes an optimal offline algorithm, and a randomized online algorithm $R$ is strictly $c$-competitive on $\mathcal{I}$ if $\mathbb{E}[\cost(I, R)] \le c \cdot \cost(I, \Opt)$ for every $I \in \mathcal{I}$. The \emph{competitive ratio} of an algorithm on $\mathcal{I}$ is the infimum over all $c$ such that it is strictly $c$-competitive on $\mathcal{I}$. We state all our results for the class of $\alpha$-instances with fixed $\alpha$ and $k$ defined below. This class is finite, so an additive constant would make every online algorithm $1$-competitive; we therefore use the strict notion and simply write $c$-competitive.

In the $k$-server problem~\cite{komm_introduction_2016,frei_time-optimal_2025}, $k$ distinguishable servers are located in a metric space $(M,d)$, and requests $r_1, \dots, r_n \in M$ arrive one by one. In step $i$, every server $v$ moves from its position $v_{i-1}$ to a position $v_i$ such that some server is located at $r_i$ afterwards, where $v_0$ is the initial position of~$v$. A configuration $C_i$ is the multiset of server positions after step $i$, and $C_i(x)$ is the number of servers at $x$. In the time model, all servers move simultaneously, so the cost of step $i$ is $\max_v d(v_{i-1}, v_i)$, and the total cost is the sum over all steps.

We study the following variant. The metric space is the infinite discrete line $M = \{p_z : z \in \mathbb{Z}\}$ with $d(p_z,p_{z'}) = |z-z'|$, all $k$ servers start at a fixed point $p_o$, called the \emph{origin}, and there are exactly $n = k$ requests. Requests are \emph{cumulative}: with $\mathrm{hist}(x,i) \da |\{ j \leq i : r_j = x \}|$ denoting the number of requests to $x$ up to step $i$, a configuration $C_i$ is \emph{valid} if $C_i(x) \geq \mathrm{hist}(x,i)$ for all $x \in M$, and every configuration of an algorithm has to be valid. Moreover, a server that serves a request becomes \emph{fixed}: it stays at the requested point for the rest of the sequence. All other servers are \emph{free}. In step~$i$, a free server located at $r_i$ after the movement serves~$r_i$. Exactly $\mathrm{hist}(x,i)$ servers at $x$ are fixed after step~$i$, and since fixed servers do not move, $C_i$ is valid if and only if a free server is located at $r_i$ after the movement. Since every request fixes exactly one server and there are exactly $k$ requests, our model is equivalent to the one in which every server serves exactly one request and stays at its point afterwards.

We specify our algorithms by their configurations. We call $\max_v d(v_{i-1}, v_i)$ the \emph{time} of a movement, and we say that $C_i$ is \emph{reachable} from $C_{i-1}$ within time~$t$ if some movement from $C_{i-1}$ to $C_i$ that keeps the fixed servers in place has time at most~$t$. Each algorithm moves from $C_{i-1}$ to $C_i$ by such a movement of minimum time. Hence, exhibiting any such movement bounds the cost of step~$i$. If the exhibited movement consists of consecutive sub-movements of times $t_1, \dots, t_m$, then by the triangle inequality no server travels farther than $t_1 + \dots + t_m$, which therefore also bounds the cost of step~$i$.

We use the following notation. The number of free servers at $x$ after step $i$ is
\[
\mathrm{surplus}(x,i) \da C_i(x) - \mathrm{hist}(x,i).
\]
Let $\mathrm{maxdist}(i) \da \max_{j \leq i} d(p_o, r_j)$ for $i \in \{1, \dots, k\}$. We call $I = (r_1, \dots, r_k)$ an \emph{$\alpha$-instance} if $\mathrm{maxdist}(k) = \alpha$. We only consider $\alpha \in \natp$, since for $\alpha = 0$ all requests lie at the origin and they can be served without any movement. For $\beta \in \nat$, let 
\[
M_\beta \da \{ x \in M : d(p_o, x) \leq \beta \},
\qquad |M_\beta| = 2\beta+1.
\]
Let $H(n) \da \sum_{i=1}^{n} 1/i$ denote the $n$-th harmonic number, where $H(0) = 0$. Finally, recall that 
\[
\mu_\alpha = 1/\ln(1+\frac{1}{2\alpha}) \in (2\alpha, 2\alpha+1) \qquad \text{for } \alpha \in \natp,
\]
where the bounds follow from $x/(1+x) < \ln(1+x) < x$ for $x > 0$. Unless stated otherwise, $\alpha, k \in \natp$ are arbitrary, $I = (r_1, \dots, r_k)$ denotes an $\alpha$-instance, and $\algA$ denotes a deterministic online algorithm.
\newpage

\begin{lemma}[Optimal Offline Cost]
\label{lemma:opt_cost}
Every $\alpha$-instance $I$ has $\cost(I, \Opt) = \alpha$.
\end{lemma}
\begin{proof}
Some request has distance $\alpha$ from the origin, and since all servers start there, every algorithm pays at least $\alpha$. Conversely, an offline algorithm can move all servers to their final positions $r_1, \dots, r_k$ in the first step, which costs $\max_j d(p_o, r_j) = \alpha$. Its configurations satisfy $C_i(x) = \mathrm{hist}(x,k) \geq \mathrm{hist}(x,i)$ for all $i$ and $x$, so they are all valid.
\end{proof}

\section{Deterministic Lower Bound}\label{sec:lower}
\paragraph{The adversary.}
Let $\alpha \in \natp$. The adversary first requests the point $r_1 \da p_{o+\alpha}$ at distance $\alpha$ from the origin, which forces a cost of at least $\alpha$, since all servers start at the origin. All further requests lie in $M_\alpha$. Before request $r_j$, there are $k-(j-1)$ free servers. As long as $k-(j-1) \geq |M_\alpha|$ (Phase~1), the adversary requests a point of $M_\alpha$ with minimum surplus, which eventually forces a movement once no free server remains there. Afterwards (Phase~2), it requests a point of $M_\alpha$ that maximizes the distance to the closest free server, which forces a movement of at least this distance. The pseudocode is given in Algorithm~\ref{alg:det_adversary}. Since every instance produced by the adversary is an $\alpha$-instance and our arguments never use that $\alpha$ is unknown, our lower bounds hold even if $\alpha$ is known. We analyze the two phases separately.

\begin{algorithm}[H]
\caption{Deterministic Adversary Strategy}\label{alg:det_adversary}
\begin{algorithmic}[1]
\Procedure{DeterministicAdversary}{$o, k, \alpha$}
    \State $r_1 \gets p_{o+\alpha}$
        \Comment{distance $\alpha$ from the origin}
    \For{$j = 2$ \textbf{to} $k$}
        \If{$k-(j-1) \geq |M_\alpha|$} \Comment{Phase~1}
            \State $r_j \in \arg\min_{x \in M_\alpha}\, \mathrm{surplus}(x,\, j-1)$
        \Else \Comment{Phase~2}
            \State $r_j \in \arg\max_{x \in M_\alpha}
                \!\left(\min_{\substack{y \in M \\\mathrm{surplus}(y,\,j-1)>0}} d(x, y)\right)$
        \EndIf
    \EndFor
\EndProcedure
\end{algorithmic}
\end{algorithm}

\paragraph{Phase~1.}
We first show that the adversary forces a movement after few requests, and then count these forced movements.

\begin{lemma}[Forced Movement in Phase~1]
\label{lemma:lower_max_req_phase1}
Let $j \in \natp$ with $k-j \geq |M_\alpha|$, and let 
\[
s \da \min_{x \in M_\alpha} \mathrm{surplus}(x,j).
\]
Then the adversary forces $\algA$ to move a server, at a cost of at least~$1$, within the requests $r_{j+1}, \dots, r_{j+s+1}$, where $s \leq \lfloor (k-j)/|M_\alpha| \rfloor$.
\end{lemma}
\begin{proof}
\label{proof:lower_max_req_phase1}
Since $C_j$ is valid, all surpluses are non-negative and sum up to $k-j$ over $M$, hence to at most $k-j$ over $M_\alpha$. Thus some $x \in M_\alpha$ has $\mathrm{surplus}(x,j) \leq \lfloor (k-j)/|M_\alpha| \rfloor$, which bounds $s$.
Now assume that $\algA$ does not move during the next $s+1$ requests. If Phase~2 starts during these requests, there are fewer free servers than points in $M_\alpha$, so the point requested by the adversary has distance at least $1$ to every free server, and $\algA$ has to move. Otherwise, the adversary requests a point $x$ with surplus $s$. This decreases the surplus of $x$ by one and leaves all other surpluses unchanged, so $x$ is now the unique point of minimum surplus and is requested again. After $s+1$ requests,
\[
\mathrm{hist}(x,j+s+1) = \mathrm{hist}(x,j)+s+1 = C_j(x)+1 = C_{j+s+1}(x)+1,
\]
so $C_{j+s+1}$ is not valid, a contradiction.
\end{proof} \newpage

\begin{lemma}[Deterministic Lower Bound for Phase~1]
\label{lemma:lower_bound_phase1}
For $k > |M_\alpha| = 2\alpha + 1$, the adversary described above
forces $\algA$ to pay a cost of at least
\[
\mu_\alpha \cdot \bigl(H(k+2\alpha) - H(4\alpha + 1) \bigr)-1
\]
to serve the requests $r_2, \dots, r_{j_1}$, where $j_1 \da k - 2\alpha$ and $\mu_\alpha = 1/\ln(1+1/(2\alpha))$.
\end{lemma}
\begin{proofidea}
By Lemma~\ref{lemma:lower_max_req_phase1}, whenever $b \geq |M_\alpha|$ requests are still to come, $\algA$ has to move within the next $\lfloor b/|M_\alpha| \rfloor + 1$ requests, after which at least $(1 - 1/|M_\alpha|)\, b - 1$ requests remain. Hence the number of remaining requests is multiplied by roughly $2\alpha/(2\alpha+1)$ per forced movement, so about $\log_{1+1/(2\alpha)} k = \mu_\alpha \ln k$ movements are forced before Phase~2. Bounding this recursion in closed form and estimating the logarithm by harmonic numbers yields the bound.
\end{proofidea}
\begin{proof}
\label{proof:lower_bound_phase1}
We first show by induction over $i \geq 0$: if $b_i \geq |M_\alpha|$, then the adversary forces $\algA$ to pay a cost of at least $i$ to serve the requests $r_2, \dots, r_{k-b_i}$, where
\[
\lambda \da \frac{|M_\alpha|-1}{|M_\alpha|}, \qquad b_i \da \left\lceil \lambda^i \cdot (k-1) - \sum_{t=0}^{i-1} \lambda^t \right\rceil \quad \text{for } i \geq 0.
\]

\textbf{Base case ($\boldsymbol{i=0}$).} We have $b_0 = k-1$, so $k-b_0 = 1$, and the claim states that $\algA$ pays a cost of at least $0$ to serve the empty sequence $r_2, \dots, r_1$, which is clearly true. \medskip

\textbf{Induction step.} Fix $i \geq 0$ with $b_{i+1} \geq |M_\alpha|$. Since $b_i \geq b_{i+1}$, as the expression inside the ceiling decreases in $i$, we have $b_i \geq |M_\alpha|$, so the induction hypothesis applies. By Lemma~\ref{lemma:lower_max_req_phase1} (applied with $j \da k-b_i$), $\algA$ is forced to move a server within the next $q+1$ requests $r_{k-b_i+1}, \dots, r_{k-b_i+q+1}$, where $q \da \lfloor b_i/|M_\alpha| \rfloor$, at a cost of at least $1$. As this movement occurs after request $r_{k-b_i}$, its cost is distinct from the cost counted by the induction hypothesis, so $\algA$ pays a total cost of at least $i+1$ to serve the requests $r_2, \dots, r_{(k-b_i)+q+1}$. It remains to show that $(k-b_i)+q+1 \leq k-b_{i+1}$.
We have
\begin{align*}
b_i - \left( q + 1 \right)
&\geq b_i - \frac{b_i}{|M_\alpha|} - 1 \tag{since $x \geq \left\lfloor x \right\rfloor$} \\
&= b_i \cdot \lambda - 1 \\
&\geq \left(\lambda^i  \cdot (k-1) - \sum_{t=0}^{i-1} \lambda^t \right) \cdot \lambda - 1 \tag{def $b_i$, $\lceil x \rceil \geq x$} \\
&= \lambda^{i+1}  \cdot (k-1)  - \sum_{t=0}^{i} \lambda^{t}.
\end{align*}
Since $b_i$ and $q$ are integers, $b_i-(q+1)$ is an integer that is at least the right-hand side, and since $b_{i+1}$ is by definition the smallest such integer, we get
\[
b_{i+1} = \left\lceil \lambda^{i+1}  \cdot (k-1)  - \sum_{t=0}^{i} \lambda^{t} \right\rceil \leq b_i -(q+1),
\]
and therefore $(k-b_i)+q+1 \leq k-b_{i+1}$. Hence $\algA$ pays a cost of at least $i+1$ to serve the requests $r_2, \dots, r_{k-b_{i+1}}$, and the induction statement follows for $i+1$.

We now determine how large $i$ can become while $b_i \geq |M_\alpha|$ still holds. In the first step, we use that $b_i$ is an integer and evaluate the geometric sum:
\begin{align*}
b_{i} < |M_\alpha|
&\implies \lambda^i \cdot (k-1) - \frac{1-\lambda^i}{1-\lambda} \leq |M_\alpha|-1 \tag{def.\ $b_i$} \\
&\implies \lambda^i \cdot (k+|M_\alpha|-1) \leq 2|M_\alpha|-1 \tag{$\frac{1}{1-\lambda} = |M_\alpha|$} \\
&\implies \frac{k+|M_\alpha|-1}{2|M_\alpha|-1} \leq \left(\frac{1}{\lambda}\right)^i \\
&\implies \mu_\alpha \cdot \ln\left(\frac{k+|M_\alpha|-1}{2|M_\alpha|-1}\right) \leq i \tag{$\frac{1}{\lambda} = 1 + \frac{1}{2\alpha}$, def.\ $\mu_\alpha$} \\
&\implies \mu_\alpha \cdot \left(H(k+|M_\alpha|-1) - H(2|M_\alpha|-1) \right) \leq i \tag{$*$}
\end{align*}
For $(*)$ note that $\frac{1}{l} \leq \int_{l-1}^{l} \frac{1}{x}\,dx$ for all $l \geq 2$, hence $\sum_{l=n+1}^{m}\frac{1}{l} \leq \int_n^m \frac{1}{x}\,dx = \ln\!\left(\frac{m}{n}\right)$ for all integers $m > n \geq 1$.

Therefore,
\[
i < \mu_\alpha \cdot \left(H(k+|M_\alpha|-1) - H(2|M_\alpha|-1) \right) \implies b_{i} \geq |M_\alpha|.
\]
Since $i$ is an integer, the largest $i$ for which the induction statement applies is at least
\[
\mu_\alpha \cdot \bigl(H(k+|M_\alpha|-1) - H(2|M_\alpha|-1)\bigr) - 1.
\]
This means that the adversary forces $\algA$ to pay a cost of at least
\begin{align*}
    &\mu_\alpha \cdot \big(H(k+|M_\alpha|-1) - H(2|M_\alpha|-1) \big)-1 \\
    &= \mu_\alpha \cdot \big(H(k+2\alpha) - H(4\alpha + 1) \big)-1.
\end{align*}
Finally, $b_i \geq |M_\alpha|$ implies $k-b_i \leq k-2\alpha-1 < j_1$, so the requests $r_2, \dots, r_{k-b_i}$ lie among $r_2, \dots, r_{j_1}$, which proves the lemma.
\end{proof}

\paragraph{Phase~2.}
In Phase~2, every request costs at least the distance from the requested point to the closest free server.

\begin{lemma}[Deterministic Lower Bound for Phase~2]
\label{lemma:lower_bound_phase2}
Let $2 \leq j \leq k$ with $k-j+1 < |M_\alpha|$. The adversary described above forces $\algA$ to pay a cost of at least
\[
\frac{2\alpha + 1}{2} H(k-j+1) - \frac{k-j+1}{2}
\]
to serve the requests $r_j, \dots, r_k$.
\end{lemma}
\begin{proof}
\label{proof:lower_bound_phase2}
Consider a request $r_l$ with $l \in \{j, \dots, k\}$. Before this request is served, $l-1$ servers are fixed and $k-l+1$ servers are free. Let
\[
S \da \{ x \in M : \mathrm{surplus}(x,l-1) > 0 \}.
\]
We have $|S| \leq k-l+1 < |M_\alpha|$ and $|S| \geq 1$.

\textbf{Distance to the closest free server.} Let $d$ be the distance from the requested point $r_l$ to its closest free server, i.e., to the closest point of $S$. Since the adversary requests a point of $M_\alpha$ farthest from $S$, every point of $M_\alpha$ lies within distance $d$ of some point of $S$. On the line, each point of $S$ has at most $2d+1$ points within distance $d$, so $|M_\alpha| \leq |S| \cdot (2d+1)$. Solving for $d$ and using $|S| \leq k-l+1$, we get
\[
d \geq \frac{|M_\alpha|-|S|}{2|S|} = \frac{|M_\alpha|}{2|S|} - \frac{1}{2} \geq \frac{|M_\alpha|}{2(k-l+1)} - \frac{1}{2}.
\]
Hence $\algA$ has to pay at least this cost to serve $r_l$.

Summing over all requests, we get
\begin{align*}
\sum_{l=j}^{k} \left( \frac{|M_\alpha|}{2(k-l+1)} - \frac{1}{2}\right)
&= \frac{|M_\alpha|}{2} \sum_{l=j}^{k} \frac{1}{k-l+1} - \frac{k-j+1}{2}\\
&= \frac{|M_\alpha|}{2} \sum_{h=1}^{k-j+1} \frac{1}{h} - \frac{k-j+1}{2} \tag{$h=k-l+1$} \\
&= \frac{2\alpha + 1}{2} H(k-j+1) - \frac{k-j+1}{2}.
\end{align*}
\end{proof}

\paragraph{Competitive ratio.}
Combining the Phase~1 and Phase~2 bounds with the cost $\alpha$ of the first request, and dividing by $\cost(I,\Opt)=\alpha$ (Lemma~\ref{lemma:opt_cost}), yields the following two theorems on the competitive ratio; for $k \leq |M_\alpha|$, Phase~1 is empty.

\begin{theorem}[Deterministic Lower Bound for $k \leq 2\alpha+1$]
\label{theo:lower_bound_ratio_small_k}
For $k \leq |M_\alpha| = 2\alpha+1$ and every deterministic online algorithm $\algA$, there exists an $\alpha$-instance $I$ with
\[
\frac{\cost(I,\algA)}{\cost(I,\Opt)} \geq 1 + \frac{2\alpha+1}{2\alpha} H(k-1) - \frac{k-1}{2\alpha} \geq \frac{2\alpha+1}{2\alpha} H(k-1).
\]
\end{theorem}
\begin{proof}
\label{proof:lower_bound_ratio_small_k}

To serve the first request, $\algA$ has to pay at least $\alpha$ since the distance from the origin to the point the adversary requests at first is exactly $\alpha$. For $k = 1$, this already proves the claim, since $H(0) = 0$. So let $k \geq 2$.
Since $k-1 < |M_\alpha|$, Lemma~\ref{lemma:lower_bound_phase2} applies with $j=2$, giving a cost of at least $\frac{2\alpha+1}{2} H(k-1) - \frac{k-1}{2}$ for $r_2$ to $r_k$. Together with $r_1$, the total cost is at least $\alpha + \frac{2\alpha+1}{2} H(k-1) - \frac{k-1}{2}$.
Dividing by $\cost(I,\Opt)=\alpha$ (Lemma~\ref{lemma:opt_cost}) gives
\[
\frac{\cost(I,\algA)}{\cost(I,\Opt)} \geq 1 + \frac{2\alpha+1}{2\alpha} H(k-1) - \frac{k-1}{2\alpha}  \geq \frac{2\alpha+1}{2\alpha} H(k-1)
\]
where the last step follows since $k \leq |M_\alpha|= 2\alpha + 1$ implies $\frac{k-1}{2\alpha} \leq 1$.
\end{proof}

\begin{theorem}[Deterministic Lower Bound for $k > 2\alpha+1$]
\label{theo:lower_bound_ratio_big_k}
For $k > |M_\alpha| = 2\alpha+1$ and every deterministic online algorithm $\algA$, there exists an $\alpha$-instance $I$ with
\[
\frac{\cost(I,\algA)}{\cost(I,\Opt)} \geq \frac{\mu_\alpha}{\alpha} \bigl(H(k+2\alpha) - H(4\alpha + 1)\bigr) + H(2\alpha) - \frac{1}{\alpha},
\]
where $\mu_\alpha = 1/\ln(1+1/(2\alpha))$.
\end{theorem}
\begin{proof}
\label{proof:lower_bound_ratio_big_k}

To serve the first request, $\algA$ pays at least $\alpha$. Since $k-1 \geq |M_\alpha|$, Lemma~\ref{lemma:lower_bound_phase1} gives a Phase~1 cost of at least $\mu_\alpha \bigl(H(k+2\alpha) - H(4\alpha + 1)\bigr)-1$.
At the start of Phase~2, exactly $|M_\alpha|-1$ requests remain, so Lemma~\ref{lemma:lower_bound_phase2} (with $k-j+1=|M_\alpha|-1$) gives a Phase~2 cost of at least
\[
\frac{|M_\alpha|}{2} H(|M_\alpha|-1) - \frac{|M_\alpha|-1}{2} = \frac{2\alpha + 1}{2} H(|M_\alpha|-1) - \alpha \geq \alpha (H(2\alpha) - 1).
\]

Combining the bounds yields a total cost of at least
\begin{align*}
&\alpha + \mu_\alpha \left(H(k+2\alpha) - H(4\alpha + 1) \right)-1 + \alpha (H(2\alpha) - 1) \\
&= \mu_\alpha \left(H(k+2\alpha) - H(4\alpha + 1) \right) + \alpha H(2\alpha) -1.
\end{align*}
Dividing by $\cost(I,\Opt)=\alpha$ (Lemma~\ref{lemma:opt_cost}) gives
\[
\frac{\cost(I,\algA)}{\cost(I,\Opt)} \geq \frac{\mu_\alpha}{\alpha} \left(H(k+2\alpha) - H(4\alpha + 1)\right) + H(2\alpha) - \frac{1}{\alpha}.
\]
\end{proof}

\section{Deterministic Upper Bound}\label{sec:upper_bound}

We first assume that $\alpha$ is known; Section~\ref{sec:upper_unknown} removes this assumption.

\subsection[Known \texorpdfstring{$\alpha$}{alpha}]{Known \texorpdfstring{$\boldsymbol{\alpha}$}{alpha}}\label{sec:upper_known}

\paragraph{The algorithm.}
$\algone$ keeps the free servers evenly distributed as follows.

\begin{definition}[Evenly Distributed]
\label{def:evenly_distributed}
Let $h \colon M \to \nat$ be a request history, i.e., $h(x)$ is the number of requests to $x$ so far, and let $f \da k - \sum_{x \in M} h(x)$ be the corresponding number of free servers. A configuration $C$ is \emph{evenly distributed} on a contiguous set of points $S=\{p_z, p_{z+1}, \dots, p_{z+m-1}\}$ for $m \in \natp$ \emph{with respect to $h$} if
\begin{description}
    \item[Case $\boldsymbol{f > |S|}$:]
    all $k$ servers are placed within $S$ and every point in $S$ holds at least $s \da \lfloor f/|S| \rfloor$ free servers, i.e., $\sum_{y \in S} C(y) = k$ and $C(y) - h(y) \geq s$ for all $y \in S$;
    \item[Case $\boldsymbol{f = 0}$:] $C=h$.     
    \item[Case $\boldsymbol{1 \leq f \leq |S|}$:] for every point $y \in S$,
    \[
    C(y) = h(y) + \mathbbm{1}[y \in \{p_{\mathit{pos}_1}, \dots, p_{\mathit{pos}_{f}}\}],
    \]
    where $\mathit{pos}_t \da z + \lfloor (2t-1)|S|/(2f) \rfloor$ for $t \in \{1, \dots, f\}$.
\end{description}
We say that the servers are evenly distributed on $S$ \emph{after step $i$} if $C_i$ is evenly distributed on $S$ with respect to $\mathrm{hist}(\cdot, i)$; in this case, $f = k-i$.
\end{definition}

$\algone$ serves $r_1$ and distributes all servers evenly on $M_\alpha$. Afterwards, while there are at least $|M_\alpha|$ free servers (Phase~1), it redistributes them evenly only every $\lfloor f/|M_\alpha| \rfloor$ steps, where $f$ is the number of free servers after the redistribution, and once there are fewer (Phase~2), it redistributes them after every request. $\algone$ redistributes after a predetermined number of steps, even if the next request could be served without movement, so the redistribution schedule does not depend on the input. A suitable configuration always exists (Lemma~\ref{lemma:upper_redis_exists_phase1}); if several qualify, $\algone$ picks one by an arbitrary but fixed rule. The pseudocode of $\algone$ is given below.

\begin{algorithm}[H]
\caption{$\algone$: Even Distribution Online Algorithm for Known $\alpha$}
\begin{algorithmic}[1]
\Procedure{$\algone$}{$k, \alpha$} \Comment{requests $r_1, \dots, r_k$ arrive online}
    \State $C_0 \gets \{p_o, \dots, p_o\}$; \quad $m \gets 2\alpha + 1$
    \State \textbf{receive} $r_1$
    \State $C_1 \gets$ a configuration evenly distributed on $M_\alpha$ after step~$1$
        \Comment{in time $\alpha$}
    \State $\mathit{steps} \gets \max\left(\left\lfloor \frac{k-1}{m} \right\rfloor - 1,\, 0\right)$
    \For{$i = 2$ \textbf{to} $k$}
        \State \textbf{receive} $r_i$; \quad $\mathit{free} \gets k - i$
        \If{$\mathit{steps} > 0$} \Comment{Phase~1 -- no redistribution}
            \State $C_i \gets C_{i-1}$; \quad $\mathit{steps} \gets \mathit{steps} - 1$
            \Comment{valid by Lemma~\ref{lemma:upper_validity_phase1}}
        \ElsIf{$\mathit{free} \geq m$} \Comment{Phase~1 -- redistribution}
            \State $s \gets \left\lfloor \frac{\mathit{free}}{m} \right\rfloor$; \quad $\mathit{steps} \gets s - 1$
            \State $C_i \gets$ a configuration evenly distributed on $M_\alpha$ after step~$i$
            \Statex \hspace{4em} and reachable from $C_{i-1}$ within time $1$
            \Comment{Lemma~\ref{lemma:upper_redis_exists_phase1}}
        \Else \Comment{Phase~2}
            \State $C_i \gets$ a configuration evenly distributed on $M_\alpha$ after step~$i$
        \EndIf
        \Statex \Comment{$C_i$ is valid w.r.t.\ $\mathrm{hist}(\cdot,i)$, so a free server serves $r_i$}
    \EndFor
\EndProcedure
\end{algorithmic}
\end{algorithm}

Since $\algone$ counts the free servers after serving a request and the adversary before, the phases of $\algone$ are shifted by one, so our upper bounds distinguish $k-1 \leq |M_\alpha|$ and $k-1 > |M_\alpha|$.

\paragraph{Phase~1.}

In Phase~1, $\algone$ redistributes only every few steps. We first characterize when servers can be redistributed from one configuration to another in time~$1$.

\begin{lemma}[Characterization of Time-$1$ Redistributions]
\label{lemma:upper_time1_shift}
Let $W$ be a window of $m \geq 1$ consecutive points $p_z, \dots, p_{z+m-1}$ on the line, let $h\colon M \to \nat$ be a request history, and let $C, C'$ be two configurations that are valid with respect to $h$, i.e., $C(y) \geq h(y)$ and $C'(y) \geq h(y)$ for all $y \in M$. Let $\sigma(y) \da C(y) - h(y)$ and $\sigma'(y) \da C'(y) - h(y)$ for $y \in M$, and for $l \in \{0, \dots, m\}$ let
\[
P(l) \da \sum_{u=0}^{l-1} \sigma(p_{z+u}) \quad \text{and} \quad P'(l) \da \sum_{u=0}^{l-1} \sigma'(p_{z+u}).
\]
Assume that
\begin{align*}
    \sum_{y \in W} C(y) = \sum_{y \in W} C'(y) = k. \tag{$1$}
\end{align*}
Then $C'$ is reachable from $C$ within time~$1$, where the $h(y)$ servers at every point $y$ are fixed, if and only if
\begin{align*}
    P(l-1) \leq P'(l) \leq P(l+1) \quad \text{for all } l \in \{1, \dots, m-1\}. \tag{$2$}
\end{align*}
\end{lemma}
\begin{proof}
\label{proof:upper_time1_shift}
Fix $m \geq 1$, a window $W = \{p_z,\dots,p_{z+m-1}\} \subseteq M$, a function $h$, and two configurations $C, C'$ that are valid with respect to $h$ and satisfy condition~$(1)$. By~$(1)$, $C$ and $C'$ have no servers outside $W$, so validity gives $h(y) = 0$ and hence $\sigma(y) = \sigma'(y) = 0$ for all $y \notin W$. In particular,
\[
P(m) = \sum_{y \in W} \bigl(C(y) - h(y)\bigr) = k - \sum_{y \in W} h(y) = \sum_{y \in W} \bigl(C'(y) - h(y)\bigr) = P'(m).
\]

\textbf{Necessity.} Assume that $C'$ is reachable from $C$ within time~$1$, consider a corresponding movement, and fix $l \in \{1, \dots, m-1\}$. All servers lie within $W$ before and after the movement, and every free server moves by distance at most~$1$. Hence each of the $P(l-1)$ free servers starting on $p_z, \dots, p_{z+l-2}$ ends on one of the points $p_z, \dots, p_{z+l-1}$, so $P'(l) \geq P(l-1)$. Conversely, each of the $P'(l)$ free servers ending on $p_z, \dots, p_{z+l-1}$ starts on one of the points $p_z, \dots, p_{z+l}$, so $P'(l) \leq P(l+1)$. Thus condition~$(2)$ holds.

\textbf{Sufficiency.} Assume condition~$(2)$. Since $P(m) = P'(m)$, both configurations have the same number of free servers. We number the free servers of $C$ and of $C'$ from left to right, where servers on the same point are numbered in arbitrary order. Then we move the $n$-th free server of $C$ to the position of the $n$-th free server of $C'$ for every $n$, while all fixed servers stay. Afterwards, every point $y$ holds $h(y)$ fixed and $\sigma'(y)$ free servers, so the resulting configuration is $C'$. It remains to show that every free server moves by distance at most~$1$.
Fix $n$, and let the $n$-th free server of $C$ lie on the $a$-th point $p_{z+a-1}$ of $W$ and the $n$-th free server of $C'$ on the $b$-th point $p_{z+b-1}$. Then $a$ is the smallest index with $P(a) \geq n$ and $b$ is the smallest index with $P'(b) \geq n$. We show that $|a-b| \leq 1$.
Suppose that $b \leq a-2$. Then $b \leq m-2$, so the right inequality of~$(2)$ for $l = b$ gives $P(b+1) \geq P'(b) \geq n$, i.e., the first $b+1$ points already hold $n$ free servers of $C$. Hence $a \leq b+1$, a contradiction. Symmetrically, suppose that $a \leq b-2$. Then $a+1 \leq m-1$, so the left inequality of~$(2)$ for $l = a+1$ gives $P'(a+1) \geq P(a) \geq n$, and hence $b \leq a+1$, a contradiction.

Thus every free server moves by distance at most~$1$, and since all moves can be performed simultaneously, $C'$ is reachable from $C$ within time~$1$.
\end{proof}

Using this characterization, we now show that $\algone$ only produces valid configurations in Phase~1 and that every redistribution in Phase~1 can be done in time~$1$. Throughout, let $s_t \da \lfloor (k-t)/|M_\alpha| \rfloor$ for $t \in \{1, \dots, k\}$ and $\mathrm{leftover}(y,t) \da \mathrm{surplus}(y,t) - s_t$, which is non-negative for all $y \in M_\alpha$ whenever $\algone$ redistributes in iteration~$t$.

 After a redistribution in iteration $j$, where the initial distribution counts as a redistribution in iteration~$1$, $\algone$ has $\mathit{steps} = \max(s_j - 1, 0)$ and decrements it once per iteration. Hence it does not move in the iterations $j+1, \dots, j+s_j-1$ and redistributes next in iteration $j + \max\{1, s_j\}$. We refer to this as the \emph{redistribution schedule}.

\begin{lemma}[Validity of Configurations in Phase~1]
\label{lemma:upper_validity_phase1}
For every $i \in \natp$ with $k-i \geq |M_{\alpha}|$, the configuration $C_i$ of $\algone$ is valid, i.e.,
\[
C_i(x) \geq \mathrm{hist}(x, i) \quad \text{for all } x \in M.
\]
\end{lemma}
\begin{proof}
\label{proof:upper_validity_phase1}

Since $I$ is an $\alpha$-instance, no point will be requested outside of $M_\alpha$, so
\[
\mathrm{hist}(x, i) = 0 \quad \text{for all } x \notin M_\alpha.
\]
It remains to show $C_i(x) \geq \mathrm{hist}(x, i)$ for all $x \in M_\alpha$. Let $j \leq i$ be the step of the most recent redistribution, which exists since $\algone$ performs an initial distribution. The servers are evenly distributed at step $j$, and by the redistribution schedule, $i-j \leq s_j$, so
\begin{align*}
    C_i(x) &= C_j(x) \tag{no movement between $j$ and $i$} \\
    &\geq \mathrm{hist}(x, j) + s_j  \tag{evenly distributed at step $j$}\\
    &\geq \mathrm{hist}(x,i) \tag{$\mathrm{hist}(x,i) - \mathrm{hist}(x, j) \leq i-j \leq s_j$ }.
\end{align*}
Hence $C_i$ is valid.
\end{proof}

\begin{lemma}[Existence of a Time-$1$ Redistribution in Phase~1]
\label{lemma:upper_redis_exists_phase1}
Let $i \geq 2$ with $k-i \geq |M_{\alpha}|$. If $\algone$ performs a redistribution at step $i$, then there exists a configuration $C_i$ that is evenly distributed on $M_\alpha$ and reachable from $C_{i-1}$ within time~$1$.
\end{lemma}
\begin{proof}
\label{proof:upper_redis_exists_phase1}

Let $j < i$ be the most recent redistribution, which exists since $\algone$ redistributes in the first step and $i \geq 2$. Since $k-j > k-i \geq |M_\alpha|$, we have $s_j \geq 1$, so the redistribution schedule gives $i - j = s_j$.
In the following, we only use $i - j \leq s_j$, so the argument also applies whenever at most $s_j$ steps have passed since an evenly distributed configuration. Moreover, $C_j$ is evenly distributed on $M_\alpha$ with surplus $s_j$:
\[
    \mathrm{surplus}(x, j) \geq s_j \quad \text{for all } x \in M_\alpha.
\]

We construct $C_i$ satisfying the two conditions of Lemma~\ref{lemma:upper_time1_shift} with $W \da M_\alpha$, $C \da C_{i-1}$ and $C' \da C_i$. Since the redistribution takes place before $r_i$ is served, the relevant history is $\mathrm{hist}(\cdot, i-1)$, so the two surplus functions of the lemma are
\begin{align*}
&\sigma(y) = C_{i-1}(y) - \mathrm{hist}(y,i-1) = \mathrm{surplus}(y,i-1) \quad\text{and}\\
&\sigma'(y) = C_i(y) - \mathrm{hist}(y,i-1) = \mathrm{surplus}(y,i) + \mathbbm{1}[y = r_i].
\end{align*}
In particular, $\sum_{y \in V'} \sigma'(y) \leq \sum_{y \in V'} \mathrm{surplus}(y,i) + 1$ for all $V' \subseteq M_\alpha$.
Both configurations are valid with respect to $\mathrm{hist}(\cdot, i-1)$: $C_{i-1}$ is valid by Lemma~\ref{lemma:upper_validity_phase1}, and $C_i$ is evenly distributed on $M_\alpha$ with respect to $\mathrm{hist}(\cdot, i)$, so $C_i(y) \geq \mathrm{hist}(y,i) \geq \mathrm{hist}(y,i-1)$ for all $y \in M_\alpha$, while no point outside $M_\alpha$ is ever requested.

\textbf{Condition~(1).}
Since $\algone$ only redistributes onto $M_\alpha$, all servers remain within $M_\alpha$, so
\[
    \sum_{y \in W} C_{i-1}(y) = \sum_{y \in W} C_{i}(y) = k.
\]

\textbf{Condition~(2).}
We show that for all $d \in \{1, \dots, |M_\alpha|-1\}$, every set $V \subseteq W$ of $d+1$ consecutive points, and every set $V' \subset V$ of $d$ consecutive points,
\[
    \sum_{y \in V} \sigma(y) \geq \sum_{y \in V'} \sigma'(y).
\]
Taking $V$ as the first $d+1$ points of $W$ and $V'$ as its first $d$ points yields the right inequality $P'(l) \leq P(l+1)$ of condition~$(2)$ of Lemma~\ref{lemma:upper_time1_shift} for $l = d$, and taking $V$ as the last $d+1$ points of $W$ and $V'$ as its last $d$ points yields, together with $\sum_{y \in W}\sigma(y) = \sum_{y \in W}\sigma'(y)$, the left inequality $P(l-1) \leq P'(l)$ for $l = |M_\alpha|-d$.

Since no server moved between iterations $j$ and $i-1$, we have $C_{j}(y) = C_{i-1}(y)$ for all $y \in V$.

Hence $C_{i-1}(y) = \mathrm{hist}(y,j) + s_j + \mathrm{leftover}(y,j)$ for all $y \in V$. Since at most $i-1-j$ requests arrive between steps $j$ and $i-1$, and $i-j \leq s_j$, we get
\begin{align*}
\sum_{y \in V} \sigma(y)
&= (d+1) s_j - \sum_{y \in V} \bigl(\mathrm{hist}(y, i-1) - \mathrm{hist}(y, j)\bigr) + \sum_{y \in V} \mathrm{leftover}(y, j) \\
&\geq (d+1) s_j - (i-1-j) + \sum_{y \in V} \mathrm{leftover}(y, j) \\
&\geq d \cdot s_j + 1 + \sum_{y \in V} \mathrm{leftover}(y, j).
\end{align*}
On the other hand, as noted above, and since $|V'| = d$,
\[
\sum_{y \in V'} \sigma'(y) \leq \sum_{y \in V'} \mathrm{surplus}(y,i) + 1 = d \cdot s_i + \sum_{y \in V'} \mathrm{leftover}(y,i) + 1.
\]
Hence it suffices to show
\[
d \cdot s_i + \sum_{y \in V'} \mathrm{leftover}(y,i) \leq d \cdot s_j + \sum_{y \in V} \mathrm{leftover}(y,j). \tag{$\ast$}
\]
The total number of leftover servers in iteration $i$ is $k-i-s_i|M_\alpha| \leq |M_\alpha|-1$, and we choose where to place them in $C_i$ depending on $s_i$.

\begin{description}
    \item[Case $\boldsymbol{s_i = s_j}$:] Since $k-i-s_i|M_\alpha| < k-j-s_j|M_\alpha|$, we can place the leftover servers such that $\mathrm{leftover}(y,i) \leq \mathrm{leftover}(y,j)$ for every $y \in M_\alpha$. Since $V' \subset V$ and $\mathrm{leftover}(y,j) \geq 0$, this implies~$(\ast)$.
    \item[Case $\boldsymbol{s_i < s_j}$:] We place the leftover servers such that every point has at most one. Then $\sum_{y \in V'} \mathrm{leftover}(y,i) \leq d$, so the left-hand side of~$(\ast)$ is at most $d \cdot (s_i+1) \leq d \cdot s_j$, which implies~$(\ast)$ since $\mathrm{leftover}(y,j) \geq 0$.
\end{description}

Both conditions of Lemma~\ref{lemma:upper_time1_shift} are therefore satisfied, so $C_i$ is reachable from $C_{i-1}$ within time~$1$.
\end{proof}

It therefore remains to count the redistributions.

\begin{lemma}[Cost of $\algone$ in Phase~1]
\label{lemma:upper_bound_phase1}
For $k - 1 > |M_\alpha| = 2\alpha + 1$, the algorithm $\algone$ serves the requests $r_2, \dots, r_{j_1'}$ with $j_1' \da k - |M_\alpha|= k - 2\alpha - 1$ within total time at most
\[
\varphi_\alpha(k) \da
\begin{cases}
k - 2\alpha - 2 & \text{if } k \leq 4\alpha+2, \\[2pt]
\mu_\alpha \ln\!\Bigl(\dfrac{k-2\alpha-1}{2\alpha+2}\Bigr) + 2\alpha + 1 & \text{if } k \geq 4\alpha+3,
\end{cases}
\]
where $\mu_\alpha = 1/\ln(1+1/(2\alpha))$. Moreover, the last redistribution of $\algone$ in Phase~1 takes place at $r_{j_1'}$.
\end{lemma}
\begin{proofidea}
After a redistribution with $b$ free servers, the next one follows $\lfloor b/|M_\alpha| \rfloor$ steps later, so $b-2\alpha$ is multiplied by at most $2\alpha/(2\alpha+1)$. For $k \geq 4\alpha+3$, this gives at most $\mu_\alpha \ln\bigl(\frac{k-2\alpha-1}{2\alpha+2}\bigr)$ redistributions after which $b \geq 2|M_\alpha|$, matching the leading term of Lemma~\ref{lemma:lower_bound_phase1}. Once $b < 2|M_\alpha|$, it decreases by exactly~$1$ per redistribution until it reaches $|M_\alpha|$ at $r_{j_1'}$. Including the redistribution that first yields $b < 2|M_\alpha|$, this adds at most $2\alpha+1$. For $k \leq 4\alpha+2$, $b < 2|M_\alpha|$ already holds after the initial distribution, so only the second part occurs.
\end{proofidea}
\begin{proof}
\label{proof:upper_bound_phase1}
By Lemma~\ref{lemma:upper_validity_phase1}, every configuration $C_i$ with $k-i \geq |M_\alpha|$ is valid. Between two redistributions, $\algone$ does not move, and by Lemma~\ref{lemma:upper_redis_exists_phase1}, every redistribution in Phase~1 costs at most~$1$. Hence it suffices to count the redistributions of $\algone$ at the requests $r_2, \dots, r_{j_1'}$.

Let $b_0 \da k-1$ and $b_{i+1} \da b_i - \lfloor b_i/|M_\alpha| \rfloor$ for $i \geq 0$. The initial distribution takes place at request $r_1 = r_{k-b_0}$. If $\algone$ redistributes at request $r_{k-b_i}$ with $b_i \geq |M_\alpha|$, then by the redistribution schedule above, its next redistribution is at request $r_{k-b_i+\lfloor b_i/|M_\alpha| \rfloor} = r_{k-b_{i+1}}$, which lies in Phase~1 if $b_{i+1} \geq |M_\alpha|$ and in Phase~2 otherwise. Hence the Phase~1 redistributions at the requests $r_2, \dots, r_{j_1'}$ are exactly those at the requests $r_{k-b_i}$ with $i \geq 1$ and $b_i \geq |M_\alpha|$, so it suffices to count these indices~$i$. While $b_i \geq |M_\alpha|$, we have $\lfloor b_i/|M_\alpha| \rfloor \geq 1$, so the sequence is strictly decreasing.

Moreover, if $|M_\alpha| \leq b_i < 2|M_\alpha|$, then $\lfloor b_i/|M_\alpha| \rfloor = 1$, so $b_{i+1} = b_i - 1$, and if $b_i \geq 2|M_\alpha|$, then
\[
b_{i+1} \geq b_i - \frac{b_i}{|M_\alpha|} = b_i \cdot \frac{|M_\alpha|-1}{|M_\alpha|} \geq 2(|M_\alpha|-1) \geq |M_\alpha|,
\]
since $|M_\alpha| \geq 3$. Hence, from a value greater than $|M_\alpha|$, the sequence never drops below $|M_\alpha|$, and once it is below $2|M_\alpha|$, it decreases in steps of~$1$. In particular, it attains the value $|M_\alpha|$, so the last redistribution in Phase~1 takes place at request $r_{k-|M_\alpha|} = r_{j_1'}$, which proves the second claim of the lemma. \smallskip

\textbf{Case $\boldsymbol{k \leq 4\alpha+2}$.} Then $b_0 = k-1 \leq 2|M_\alpha|-1$, so $b_i = b_0 - i$ as long as $b_i \geq |M_\alpha|$. Hence the indices $i \geq 1$ with $b_i \geq |M_\alpha|$ are $1, \dots, b_0 - |M_\alpha|$, and there are $b_0 - |M_\alpha| = k - 2\alpha - 2 = \varphi_\alpha(k)$ of them. \smallskip

\textbf{Case $\boldsymbol{k \geq 4\alpha+3}$.} Then $b_0 = k-1 \geq 2|M_\alpha|$. Let $t \geq 1$ be the smallest index with $b_t < 2|M_\alpha|$. We count the indices $i \geq 1$ with $b_i \geq 2|M_\alpha|$, which are $1, \dots, t-1$, and those with $|M_\alpha| \leq b_i < 2|M_\alpha|$ separately.

For the first ones, we use $\lfloor b_i/|M_\alpha| \rfloor \geq (b_i - |M_\alpha| + 1)/|M_\alpha| = (b_i - 2\alpha)/(2\alpha+1)$, which gives
\[
b_{i+1} - 2\alpha = b_i - 2\alpha - \left\lfloor \frac{b_i}{|M_\alpha|} \right\rfloor \leq \frac{2\alpha}{2\alpha+1} \cdot (b_i - 2\alpha).
\]
Applying this $t-1$ times and using $b_{t-1} - 2\alpha \geq 2|M_\alpha| - 2\alpha = 2\alpha+2$, we get
\[
2\alpha+2 \leq \Bigl(\frac{2\alpha}{2\alpha+1}\Bigr)^{t-1} (b_0 - 2\alpha) = \Bigl(\frac{2\alpha}{2\alpha+1}\Bigr)^{t-1} (k-2\alpha-1),
\]
so $t - 1 \leq \mu_\alpha \ln\bigl(\frac{k-2\alpha-1}{2\alpha+2}\bigr)$, since $\mu_\alpha = 1/\ln(1+\frac{1}{2\alpha})$.

For the second ones, we have $b_t < 2|M_\alpha|$ by the choice of $t$, and $b_t \geq |M_\alpha|$ since $b_{t-1} \geq 2|M_\alpha|$. As the sequence decreases in steps of~$1$ from $b_t$ on, the indices $i \geq t$ with $b_i \geq |M_\alpha|$ are exactly those with $b_i \in \{b_t, b_t - 1, \dots, |M_\alpha|\}$, so there are $b_t - |M_\alpha| + 1 \leq |M_\alpha| = 2\alpha+1$ of them.

Adding both, $\algone$ redistributes at most $\mu_\alpha \ln\bigl(\frac{k-2\alpha-1}{2\alpha+2}\bigr) + 2\alpha + 1 = \varphi_\alpha(k)$ times at the requests $r_2, \dots, r_{j_1'}$.

In both cases, since every redistribution costs at most~$1$, $\algone$ serves the requests $r_2, \dots, r_{j_1'}$ within total time at most $\varphi_\alpha(k)$.
\end{proof}

\paragraph{Phase~2.}
In Phase~2, $\algone$ keeps the free servers evenly distributed, so that every point of $M_\alpha$ is close to some free server.

\begin{lemma}[Cost of $\algone$ in Phase~2]
\label{lemma:upper_bound_phase2}
For $k \geq 2$, let $j \da k - |M_\alpha| + 1$ if $k - 1 > |M_\alpha|$ and $j \da 2$ otherwise. Then $\algone$ serves the requests $r_j, \dots, r_k$ within total time at most
\[
(2\alpha + 1) \cdot H(k-j+1) + \alpha.
\]
\end{lemma}
\begin{proofidea}
When Phase~2 begins, the servers are evenly distributed on $M_\alpha$, and $\algone$ restores this distribution after every request; serving a request and redistributing the remaining free servers happen simultaneously. With $f \da k-l+1$ free servers before $r_l$, neighboring free servers are at most $|M_\alpha|/f + 1$ apart and the outermost ones are at most $|M_\alpha|/(2f)$ away from the ends of $M_\alpha$, so serving $r_l$ costs at most $\lfloor |M_\alpha|/(2f) \rfloor + 1$. After the serving server is removed, comparing the old and new positions shows that every remaining server moves by less than $|M_\alpha|/f + 1$, so $r_l$ costs at most $\lfloor |M_\alpha|/f \rfloor + 1$. The last request needs no redistribution and costs at most $\lfloor |M_\alpha|/2 \rfloor + 1 = \alpha + 1$. Summing over all requests yields the bound, since the $k-j+1 \leq |M_\alpha|$ additive ones are absorbed by the first term of the harmonic sum.
\end{proofidea}
\begin{proof}
\label{proof:upper_bound_phase2}
Let $z_{\min} \da o - \alpha$, so that $M_\alpha = \{p_{z_{\min}}, \dots, p_{z_{\min}+|M_\alpha|-1}\}$. We first show, by case distinction on $k$, that the servers are evenly distributed before every request in Phase~2:
\begin{description}
    \item[Case $\boldsymbol{k-1 \leq |M_\alpha|}$:]
    Since $k-1 \leq |M_\alpha|$, the initial distribution performed by $\algone$ at iteration $1$ places the servers as in Definition~\ref{def:evenly_distributed}.

    \item[Case $\boldsymbol{k-1 > |M_\alpha|}$:]
    Then request $r_j$ is the first request where $\algone$ entered the Phase~2 branch and the last redistribution was in the Phase~1 branch. By Lemma~\ref{lemma:upper_bound_phase1}, this last redistribution takes place at the request $r_i$ with $k-i = |M_\alpha|$, i.e., $i = j-1$. At this redistribution, $f = k-i = |M_\alpha|$ servers are free, and $\algone$ chooses a configuration that is evenly distributed on $M_\alpha$.     
    By Definition~\ref{def:evenly_distributed}, we then have $\mathit{pos}_t = z_{\min} + t - 1$, i.e.,
    \[
    C_i(x) = \mathrm{hist}(x,i) + 1 \quad \text{for all } x \in M_\alpha,
    \]
    so also in this case the servers were evenly distributed before request $r_j$.
\end{description}
Since all subsequent requests always enter the Phase~2 branch, the servers are thus evenly distributed before every request $r_j, \dots, r_k$.

Fix $l \in \{j, \dots, k\}$, and assume the servers are evenly distributed after step $l-1$, i.e., the free servers sit at positions
\[
    \mathit{pos}_t = z_{\min} + \left\lfloor \frac{(2t-1)|M_\alpha|}{2(k-l+1)} \right\rfloor \quad \text{for } t \in \{1, \dots, k-l+1\}.
\]
For each request $r_l$, $\algone$ has two simultaneous movements: serving $r_l$, and redistributing the remaining $k-l$ free servers. We bound each one. \smallskip

\textbf{Distance to the closest free server.} Let $x \da |M_\alpha|/(2(k-l+1))$. The leftmost free server has distance $\mathit{pos}_1 - z_{\min} = \lfloor x \rfloor$ from the left end of $M_\alpha$, and the rightmost one has distance 
\[ (z_{\min}+|M_\alpha|-1) - \mathit{pos}_{k-l+1} = |M_\alpha|-1-\lfloor |M_\alpha| - x \rfloor = \lceil x \rceil - 1 \leq \lfloor x \rfloor \]
from the right end. Since $\lfloor a \rfloor - \lfloor b \rfloor \leq a-b+1$, neighboring free servers have distance ${\mathit{pos}_{t+1} - \mathit{pos}_t \leq 2x+1}$, so every point between them has distance at most $\lfloor x + \frac12 \rfloor \leq \lfloor x \rfloor + 1$ to the closer one. Hence serving $r_l$ costs at most $\lfloor x \rfloor + 1$. \newpage

\textbf{Time for redistribution.} For $l=k$, there is only one free server and no redistribution is needed. So let $l<k$. The $k-l+1$ free servers after step $l-1$ sit at positions $\mathit{pos}_t$. After one server serves $r_l$, the remaining $k-l$ servers must reach
\[
\mathit{pos}_t' = z_{\min} + \left\lfloor \frac{(2t-1)|M_\alpha|}{2(k-l)} \right\rfloor, \quad t \in \{1, \dots, k-l\}.
\]
Say the server at $\mathit{pos}_{t^\star}$ serves $r_l$. We assign the servers at $\mathit{pos}_t$, $t < t^\star$, to $\mathit{pos}_t'$, and those at $\mathit{pos}_t$, $t > t^\star$, to $\mathit{pos}_{t-1}'$, so each new position gets exactly one server.

\emph{Case $t \in \{1, \dots, t^\star-1\}$.}
Since $\mathit{pos}_t' \geq \mathit{pos}_t$, the distance is
\begin{align*}
\mathit{pos}_t' - \mathit{pos}_t
&= \left\lfloor \frac{(2t-1)|M_\alpha|}{2(k-l)} \right\rfloor - \left\lfloor \frac{(2t-1)|M_\alpha|}{2(k-l+1)} \right\rfloor \\
&\leq \frac{(2t-1)|M_\alpha|}{2(k-l)(k-l+1)} + 1 \tag{$\lfloor a \rfloor - \lfloor b \rfloor \leq a-b+1$} \\
&\leq \frac{\left(2(k-l)-1\right)|M_\alpha|}{2(k-l)(k-l+1)} + 1 \tag{$t \leq k-l$}\\
&< \frac{|M_\alpha|}{k-l+1} + 1 \tag{$\frac{2(k-l)-1}{2(k-l)} < 1$}.
\end{align*}

\emph{Case $t \in \{t^\star+1, \dots, k-l+1\}$.}
Since $(2t-1)(k-l) - (2t-3)(k-l+1) = 2(k-l)-2t+3 > 0$ for $t \leq k-l+1$, we have $\mathit{pos}_t \geq \mathit{pos}'_{t-1}$, and the distance is
\begin{align*}
\mathit{pos}_t - \mathit{pos}_{t-1}'
&= \left\lfloor \frac{(2t-1)|M_\alpha|}{2(k-l+1)} \right\rfloor - \left\lfloor \frac{(2t-3)|M_\alpha|}{2(k-l)} \right\rfloor \\
&\leq \frac{(2(k-l)-2t+3)|M_\alpha|}{2(k-l)(k-l+1)} + 1 \tag{$\lfloor a \rfloor - \lfloor b \rfloor \leq a-b+1$} \\
&\leq \frac{(2(k-l)-1)|M_\alpha|}{2(k-l)(k-l+1)} + 1 \tag{$t \geq 2$}\\
&< \frac{|M_\alpha|}{k-l+1} + 1 \tag{$\frac{2(k-l)-1}{2(k-l)} < 1$}.
\end{align*}

In both cases the distance moved is strictly less than $\frac{|M_\alpha|}{k-l+1} + 1$, and since it is an integer, at most $\lfloor|M_\alpha|/(k-l+1)\rfloor+1$. Since the movements of different servers are independent, they can be done in parallel, so redistribution costs at most $\lfloor|M_\alpha|/(k-l+1)\rfloor+1$. \smallskip

\textbf{Combining both bounds.} Both movements happen simultaneously, so request $r_l$ with $l < k$ costs at most
\[
\max\bigl(\lfloor x \rfloor + 1, \lfloor |M_\alpha|/(k-l+1) \rfloor + 1\bigr) = \lfloor |M_\alpha|/(k-l+1) \rfloor + 1,
\]
and request $r_k$, for which no redistribution is needed, costs at most $\lfloor |M_\alpha|/2 \rfloor + 1$.

Summing over all requests, we get
\begin{align*}
&\left\lfloor\frac{|M_\alpha|}{2}\right\rfloor + 1 + \sum_{l=j}^{k-1}\left( \left\lfloor\frac{|M_\alpha|}{k-l+1}\right\rfloor + 1 \right) \\
&\leq \alpha + (k-j+1) + \sum_{i=2}^{k-j+1} \frac{|M_\alpha|}{i} \tag{$i = k-l+1$, $\left\lfloor\frac{|M_\alpha|}{2}\right\rfloor = \alpha$}\\
&\leq \alpha + \left(2\alpha + 1\right) \cdot H(k-j+1). \tag{$k - j + 1 \leq |M_\alpha|$}
\end{align*}
\end{proof} \newpage

\paragraph{Total cost.}
We bound the total cost of $\algone$ by summing bounds on its individual movements. For a movement $X$ of $\algone$, let $\overline{\cost}(X)$ denote the bound the corresponding lemma gives for it. We have:
\begin{itemize}
    \item the initial distribution costs $\alpha$, since all servers move from $p_o$ to positions in $M_\alpha$;
    \item a Phase~1 redistribution costs $1$ (Lemma~\ref{lemma:upper_redis_exists_phase1});
    \item a Phase~2 movement at iteration $l < k$ costs $\lfloor |M_{\alpha}|/(k-l+1) \rfloor + 1$, and at iteration $l = k$ it costs $\lfloor |M_{\alpha}|/2 \rfloor + 1$, by the proof of Lemma~\ref{lemma:upper_bound_phase2}.
\end{itemize}
Let $\overline{\cost}(I,\algone)$ be the sum of these bounds over all movements $\algone$ performs on $I$, so that $\cost(I,\algone) \leq \overline{\cost}(I,\algone)$. The following two lemmas add up these bounds, using Lemmas~\ref{lemma:upper_bound_phase1} and~\ref{lemma:upper_bound_phase2} for Phase~1 and Phase~2, so they bound $\overline{\cost}(I,\algone)$ and hence also $\cost(I,\algone)$.

\begin{lemma}[Total Cost of $\algone$ for $k-1 \leq 2\alpha+1$]
\label{lemma:upper_total_small_k}
For $k \geq 2$ with $k-1 \leq |M_\alpha|=2\alpha + 1$, $\algone$ serves all requests within total time at most
\[
\left( 2\alpha + 1\right) \cdot \left( H(k-1) + 1\right).
\]
\end{lemma}
\begin{proof}
\label{proof:upper_bound_total_small_k}

The initial distribution costs at most $\alpha$. Since $k-1 \leq |M_\alpha|$, Lemma~\ref{lemma:upper_bound_phase2} bounds the cost of requests $r_2$ to $r_k$ by $\alpha + (2\alpha+1) H(k-1)$, which, together with the initial cost, gives the claimed bound.
\end{proof}

\begin{lemma}[Total Cost of $\algone$ for $k-1 > 2\alpha+1$]
\label{lemma:upper_total_big_k}
For $k-1 > |M_\alpha|= 2\alpha + 1$, $\algone$ serves all requests within total time at most
\[
\varphi_\alpha(k) + (2\alpha+1)H(2\alpha+1) + 2\alpha,
\]
where $\varphi_\alpha(k)$ is defined as in Lemma~\ref{lemma:upper_bound_phase1}.
\end{lemma}
\begin{proof}
\label{proof:upper_bound_total_big_k}

The initial distribution costs at most $\alpha$. By Lemma~\ref{lemma:upper_bound_phase1}, the cost for requests $r_2$ to $r_{k-|M_\alpha|}$ is at most $\varphi_\alpha(k)$. By Lemma~\ref{lemma:upper_bound_phase2}, the cost for requests $r_{k-|M_\alpha|+1}$ to $r_k$ is at most $\alpha + (2\alpha+1)H(2\alpha+1)$. Adding these three bounds proves the lemma.
\end{proof}

\subsection[Unknown \texorpdfstring{$\alpha$}{alpha}]{Unknown \texorpdfstring{$\boldsymbol{\alpha}$}{alpha}}\label{sec:upper_unknown}

\paragraph{The algorithm.}
Since $\algtwo$ does not know $\alpha$ in advance, it works on the current range $M_{\alpha'}$, where $\alpha'$ is the largest distance from the origin among the requests seen so far. As long as all requests lie within this range, $\algtwo$ behaves exactly like $\algone$ with $\alpha'$ in place of $\alpha$. If $r_1 = p_o$, then $\alpha' = 0$, and all servers stay at the origin until the first request outside it, since all earlier requests can be served without movement (see below). If a request $r_i$ lies outside the current range, $\algtwo$ first enlarges the range to $M_{\mathrm{maxdist}(i)}$, redistributes the servers evenly on it, and recomputes the number of steps until its next redistribution for the new range. The pseudocode of $\algtwo$ is given in Algorithm~\ref{alg:alg2}.

\begin{algorithm}[!t]
\caption{$\algtwo$: Even Distribution Online Algorithm for Unknown $\alpha$}\label{alg:alg2}
\begin{algorithmic}[1]
\Procedure{$\algtwo$}{$o, k$} \Comment{requests $r_1, \dots, r_k$ arrive online}
    \State $C_0 \gets \{p_o, \dots, p_o\}$
    \State \textbf{receive} $r_1$
    \State $\alpha' \gets d(p_o, r_1)$; \quad $m \gets 2\alpha' + 1$
    \State \textbf{set} $C_1$ to be a configuration
    \Statex \hspace{3.5em} that is evenly distributed on $M_{\alpha'}$ after step~$1$
        \Comment{initial distribution, in time $\alpha'$}
    \State $\mathit{steps} \gets \max\left(\left\lfloor \frac{k-1}{m} \right\rfloor - 1,\, 0\right)$
    \For{$i = 2$ \textbf{to} $k$}
        \State \textbf{receive} $r_i$; \quad $\mathit{free} \gets k - i$
        \If{$d(p_o, r_i) > \alpha'$}  \Comment{Range increase}
            \State $\gamma \gets d(p_o, r_i) - \alpha'$
            \State $\tau \gets \begin{cases}
                2 + \gamma & \text{if } \mathit{free}+1 \geq m
                             \text{ and } \mathit{free}+1 < 2\,d(p_o,r_i)+1, \\
                1 + \gamma & \text{otherwise}
            \end{cases}$
            \Statex \hspace{4em} \Comment{$2+\gamma$ only when leaving Phase~1 for Phase~2,
                Lem.~\ref{lemma:upper_redis_grow_phase1},~\ref{lemma:upper_redis_grow_phase2}}
            \State $\alpha' \gets d(p_o, r_i)$; \quad $m \gets 2 \alpha' + 1$
            \State $\hat{C} \gets$ a configuration evenly distributed on $M_{\alpha'}$
            \Comment{Def.~\ref{def:evenly_distributed}}
            \Statex \hspace{4em} w.r.t.\ $\mathrm{hist}(\cdot, i-1)$ and reachable from $C_{i-1}$ within time $\tau$
            \State $s \gets \left\lfloor \frac{\mathit{free}+1}{m} \right\rfloor$
                \Comment{$\mathit{free}+1$ since $r_i$ is served later}
            \State $\mathit{steps} \gets \max(s - 1,\, 0)$
        \Else
            \State $\hat{C} \gets C_{i-1}$
        \EndIf
        \If{$\mathit{steps} > 0$} \Comment{Phase~1 -- no redistribution}
            \State $C_i \gets \hat{C}$; \quad $\mathit{steps} \gets \mathit{steps} - 1$
        \ElsIf{$\mathit{free} \geq m$} \Comment{Phase~1 -- redistribution}
            \State $s \gets \left\lfloor \frac{\mathit{free}}{m} \right\rfloor$; \quad $\mathit{steps} \gets s - 1$
            \State $C_i \gets$ a configuration evenly distributed on $M_{\alpha'}$ after step~$i$
             \Comment{Def.~\ref{def:evenly_distributed}}
            \Statex \hspace{4em} and reachable from $\hat{C}$ within time $1$
        \Else \Comment{Phase~2}
            \State $C_i \gets$ a configuration evenly distributed on $M_{\alpha'}$ after step~$i$
            \State $\mathit{steps} \gets 0$
        \EndIf
    \EndFor
\EndProcedure
\end{algorithmic}
\end{algorithm}

\paragraph{Well-definedness.}
In the following, $\alpha'$ denotes the current range of $\algtwo$, as in the pseudocode. Note that $\alpha' = d(p_o, r_1)$ may be $0$ even though $\alpha \geq 1$. Then $M_{\alpha'} = \{p_o\}$, and until the first range increase, all requests lie at the origin and are served without any movement. All definitions and lemmas also hold for $|M_{\alpha'}| = 1$, where condition~$(2)$ of Lemma~\ref{lemma:upper_time1_shift} is vacuous. Moreover, a range increase in iteration $j$ yields a configuration $\hat{C}$ that is evenly distributed with respect to $\mathrm{hist}(\cdot, j-1)$, and if $\algtwo$ keeps $\hat{C}$, it sets $\mathit{steps} = s-1$ for $s = \lfloor (k-j+1)/|M_{\alpha'}| \rfloor$, exactly as $\algone$ after a redistribution in iteration $j-1$. We therefore count such a range increase as a redistribution in iteration $j-1$; with this convention, the proofs for $\algone$ carry over to $\algtwo$.

\begin{lemma}[Properties of $\algtwo$]
\label{lemma:alg2_properties}
In every iteration $i$ of $\algtwo$, the following holds.
\begin{enumerate}
    \item[(a)] The configurations $\hat{C}$ and $C_i$ chosen by $\algtwo$ exist.
    \item[(b)] The configuration $C_i$ is valid.
    \item[(c)] If $\algtwo$ does not redistribute, then at least $2|M_{\alpha'}|-1$ servers are free. In particular, $\algtwo$ redistributes whenever $k-i < |M_{\alpha'}|$.
    \item[(d)] In every iteration $i \geq 2$ with $k-i < |M_{\alpha'}|$, the configuration $\hat{C}$, after any range increase and immediately before the subsequent serving and redistribution step, is evenly distributed on the current range $M_{\alpha'}$ with respect to $\mathrm{hist}(\cdot,i-1)$ as in Definition~\ref{def:evenly_distributed}.
\end{enumerate}
\end{lemma} \newpage
\begin{proof}
We prove (a)--(d) by induction on $i$, i.e., when proving them for iteration $i$, we may assume them for all earlier iterations. In particular, Lemmas~\ref{lemma:upper_redis_grow_phase1} and~\ref{lemma:upper_redis_grow_phase2}, which we use in~(a), may rely on (b) and~(c) for iteration $i-1$, which is all they need from this lemma.

\textbf{(a)} For a range increase from $\alpha'$ to $d(p_o, r_i) = \alpha'+\gamma$, the configuration $\hat{C}$ must be evenly distributed on $M_{\alpha'+\gamma}$ with respect to $\mathrm{hist}(\cdot, i-1)$ and reachable from $C_{i-1}$ within time $\tau$. Before $r_i$, $\mathit{free}+1 = k-(i-1)$ servers are free. If $k-(i-1) \geq |M_{\alpha'+\gamma}|$, Lemma~\ref{lemma:upper_redis_grow_phase1} gives such a configuration within time $1+\gamma$. Otherwise, Lemma~\ref{lemma:upper_redis_grow_phase2} gives one within time $2+\gamma$ if $k-(i-1) \geq |M_{\alpha'}|$, and within time $1+\gamma$ if not. The first case in the definition of $\tau$ is exactly the case of time $2+\gamma$.

For a Phase~1 redistribution, $C_i$ must be evenly distributed on $M_{\alpha'}$ and reachable from $\hat{C}$ within time~$1$. Since $\hat{C}$ is either $C_{i-1}$ or the result of a range increase in iteration~$i$, it stems from a redistribution as in $\algone$, so the proof of Lemma~\ref{lemma:upper_redis_exists_phase1} applies with $\alpha'$ in place of $\alpha$.

For a redistribution in the Phase~2 branch, we have $k-i < |M_{\alpha'}|$, so Definition~\ref{def:evenly_distributed} defines $C_i$ explicitly. Since the pseudocode requires no time bound here, $C_i$ clearly exists.

\textbf{(b)} In the redistribution branches, $C_i$ is evenly distributed with respect to $\mathrm{hist}(\cdot, i)$ and hence valid. In the branch without redistribution, the proof of Lemma~\ref{lemma:upper_validity_phase1} applies with $\alpha'$ in place of $\alpha$: after the most recent redistribution in iteration $j$, every point of the current range holds at least $s_j$ free servers, at most $s_j$ requests arrive until the next redistribution, and all of them lie in the current range.

\textbf{(c)} Let $i$ be an iteration in which $\algtwo$ does not redistribute. We show that at least $2|M_{\alpha'}|-1$ servers are free after it. Let $j < i$ be the iteration of the most recent redistribution, where a range increase in iteration $j+1$ counts as a redistribution in iteration $j$ as noted above, and let $b \da k-j$ be the number of free servers after it. By the redistribution schedule of Section~\ref{sec:upper_known} with $\alpha'$ in place of $\alpha$, $\algtwo$ redistributes next in iteration $j + \max\{1, \lfloor b/|M_{\alpha'}| \rfloor\}$. Since $i$ lies strictly between $j$ and this iteration, we have $\lfloor b/|M_{\alpha'}| \rfloor \geq 2$, i.e., $b \geq 2|M_{\alpha'}|$, and $i \leq j + \lfloor b/|M_{\alpha'}| \rfloor - 1$. Hence, after iteration $i$,
\[
k - i \geq b - \Bigl\lfloor \frac{b}{|M_{\alpha'}|} \Bigr\rfloor + 1 \geq b\Bigl(1 - \frac{1}{|M_{\alpha'}|}\Bigr) + 1 \geq 2|M_{\alpha'}| - 1
\]
servers are free. Thus $\algtwo$ redistributes whenever $k-i < 2|M_{\alpha'}|-1$, and in particular whenever $k-i < |M_{\alpha'}|$.

\textbf{(d)} Let $i \geq 2$ and assume $k-i < |M_{\alpha'}|$, where $\alpha'$ denotes the range after any increase in iteration $i$. Immediately before serving $r_i$, there are $k-(i-1) \leq |M_{\alpha'}|$ free servers. We show that $\hat{C}$ is evenly distributed with respect to $\mathrm{hist}(\cdot,i-1)$ by distinguishing three cases.

If iteration $i$ contains a range increase, then $\hat{C}$ is evenly distributed by its construction; its existence follows from~(a).

If iteration $i$ contains no range increase and $k-(i-1) < |M_{\alpha'}|$, then the movement starts from $C_{i-1}$. By~(c), iteration $i-1$ is a Phase~2 redistribution, so $C_{i-1}$ is evenly distributed.

If iteration $i$ contains no range increase and $k-(i-1) = |M_{\alpha'}|$, then the movement starts from $C_{i-1}$. If $|M_{\alpha'}| \geq 3$, then $k-(i-1) < 2|M_{\alpha'}|-1$, so by~(c), iteration $i-1$ is a redistribution, or the initial distribution if $i = 2$, and $C_{i-1}$ is evenly distributed. If $|M_{\alpha'}| = 1$, the single free server lies on $p_o$, which is the whole range.
\end{proof}

By~(c) and~(d), $\algtwo$ redistributes in every Phase~2 iteration, starting from a configuration $\hat{C}$ with evenly distributed free servers. Hence the proof of Lemma~\ref{lemma:upper_bound_phase2} applies with $\alpha'$ in place of $\alpha$: a Phase~2 redistribution in iteration $l < k$ takes time at most $\lfloor |M_{\alpha'}|/(k-l+1) \rfloor + 1$, and in iteration $k$ at most $\lfloor |M_{\alpha'}|/2 \rfloor + 1$. Range increases are bounded separately by~$\tau$.

\paragraph{Range increases.}
Let $\gamma \da \mathrm{maxdist}(i) - \mathrm{maxdist}(i-1)$ be the increase in range caused by a request $r_i$ outside the current range.  If there are still at least $|M_{\mathrm{maxdist}(i)}|$ free servers before $r_i$ (Phase~1 with respect to the new range), the redistribution onto the larger range can be split into $1+\gamma$ sub-movements that each satisfy the conditions of Lemma~\ref{lemma:upper_time1_shift}: first on the old range, then adding one point on each side at a time. Otherwise (Phase~2 with respect to the new range), every server moves by at most $1+\gamma$, plus one additional step if the previous request was still in Phase~1 with respect to the old range. Hence every range increase takes time at most $1+\gamma$, or $2+\gamma$ if it leads from Phase~1 with respect to the old range to Phase~2 with respect to the new range (Lemmas~\ref{lemma:upper_redis_grow_phase1} and~\ref{lemma:upper_redis_grow_phase2}).

\begin{lemma}[Time of a Range Increase Ending in Phase~1]
\label{lemma:upper_redis_grow_phase1}
Let $i \in \{2, \dots, k\}$, let the range of $\algtwo$ before iteration $i$ be $\alpha' \da \mathrm{maxdist}(i-1)$, and let $\gamma \geq 0$ with $k-(i-1) \geq |M_{\alpha'+\gamma}|$.
Then there exists a configuration $\hat{C}$ that is evenly distributed on $M_{\alpha'+\gamma}$ with respect to $\mathrm{hist}(\cdot, i-1)$ and reachable from $C_{i-1}$ within time $1+\gamma$.
\end{lemma}
\begin{proof}
\label{proof:upper_redis_grow_phase1}
Let $f \da k-(i-1)$ be the number of free servers before $r_i$, and let $s^l \da \lfloor f/|M_{\alpha'+l}| \rfloor$ for $l \in \{0, \dots, \gamma\}$; note that $s^l \geq 1$ since $f \geq |M_{\alpha'+\gamma}|$. We move the servers in $\gamma+1$ consecutive sub-movements
\[
C_{i-1} \to C^0 \to C^1 \to \dots \to C^{\gamma} = \hat{C}.
\]
Each $C^l$ is evenly distributed on $M_{\alpha'+l}$ with respect to $\mathrm{hist}(\cdot, i-1)$. By the triangle inequality (Section~\ref{sec:prelim}), it suffices to bound each sub-movement by time~$1$. For $y \in M_{\alpha'+l}$, we write
\[
\mathrm{surplus}^l(y) \da C^l(y) - \mathrm{hist}(y, i-1) \quad \text{and} \quad \mathrm{leftover}^l(y) \da \mathrm{surplus}^l(y) - s^l \geq 0.
\]

\textbf{From $\boldsymbol{C_{i-1}}$ to $\boldsymbol{C^0}$.} Let $j$ be the most recent iteration before $i$ in which $\algtwo$ redistributed. As above, a range increase in an iteration $j+1 < i$ counts as a redistribution in iteration $j$. Since $f \geq |M_{\alpha'+\gamma}| \geq |M_{\alpha'}|$, this redistribution was in Phase~1. If $j = i-1$, then $C_{i-1}$ is already evenly distributed with respect to $\mathrm{hist}(\cdot, i-1)$, and we choose $C^0 \da C_{i-1}$. Otherwise, $\algtwo$ has not moved since iteration $j$, and $i - j \leq \lfloor (k-j)/|M_{\alpha'}| \rfloor$ by the redistribution schedule of Section~\ref{sec:upper_known}. These are the only properties of $\algone$ that the proof of Lemma~\ref{lemma:upper_redis_exists_phase1} uses, and $C_{i-1}$ is valid by Lemma~\ref{lemma:alg2_properties}(b). Hence the calculation in that proof applies with $\alpha'$ in place of $\alpha$, and with $C^0$ and $s^0$ in place of $C_i$ and $s_i$, so $C^0$ is reachable from $C_{i-1}$ within time~$1$. \medskip

\textbf{From $\boldsymbol{C^{l-1}}$ to $\boldsymbol{C^l}$ for $\boldsymbol{l \in \{1, \dots, \gamma\}}$.} Let $W_{\mathrm{old}} \da M_{\alpha'+l-1}$ and $W \da M_{\alpha'+l}$, so $W$ contains two new points, one at each end. These lie outside $M_{\alpha'}$, so they have never been requested and hold no server in $C^{l-1}$. We apply Lemma~\ref{lemma:upper_time1_shift} with the window $W$. Since $C^{l-1}$ and $C^l$ are evenly distributed on $W_{\mathrm{old}}$ and $W$, respectively, both are valid with respect to $\mathrm{hist}(\cdot, i-1)$, and all their servers lie within $W \supseteq W_{\mathrm{old}}$, so condition~$(1)$ holds. For condition~$(2)$, as in the proof of Lemma~\ref{lemma:upper_redis_exists_phase1}, it suffices to show for all $d \in \{1, \dots, |W|-1\}$, every set $V \subseteq W$ of $d+1$ consecutive points, and every set $V' \subset V$ of $d$ consecutive points that
\[
\sum_{y \in V} \mathrm{surplus}^{l-1}(y) \geq \sum_{y \in V'} \mathrm{surplus}^l(y).
\]
Taking $V'$ as the first or last $d$ points of $V$ then yields the two inequalities of condition~$(2)$, since both configurations have $f$ free servers in total.
If $V = W$, the left-hand side equals $f$, which is at least the right-hand side. Otherwise, $V$ contains at most one of the two new points, so at least $d$ points of $V$ lie in $W_{\mathrm{old}}$ and hold at least $s^{l-1}$ free servers each in $C^{l-1}$. Since $|W| > |W_{\mathrm{old}}|$, we have $s^l \leq s^{l-1}$, and we place the $f - s^l|W|$ leftover servers in $C^l$ as follows.

\begin{description}
    \item[Case $\boldsymbol{s^l < s^{l-1}}$:] We place at most one leftover server on every point of $W$, which is possible since $f - s^l|W| < |W|$. Then
    \[
    \sum_{y \in V} \mathrm{surplus}^{l-1}(y) \geq d \cdot s^{l-1} \geq d \cdot (s^l + 1) \geq \sum_{y \in V'} \mathrm{surplus}^l(y).
    \]
    \item[Case $\boldsymbol{s^l = s^{l-1}}$:] We place no leftover server on the two new points and at most $\mathrm{leftover}^{l-1}(y)$ on every $y \in W_{\mathrm{old}}$, which is possible since
    \[
    f - s^l|W| = f - s^{l-1}|W_{\mathrm{old}}| - 2s^l \leq f - s^{l-1}|W_{\mathrm{old}}| = \sum_{y \in W_{\mathrm{old}}} \mathrm{leftover}^{l-1}(y).
    \]
    Then
    \begin{align*}
    \sum_{y \in V} \mathrm{surplus}^{l-1}(y)
    &\geq d \cdot s^{l-1} + \sum_{y \in V \cap W_{\mathrm{old}}} \mathrm{leftover}^{l-1}(y) \\
    &\geq d \cdot s^l + \sum_{y \in V'} \mathrm{leftover}^l(y) = \sum_{y \in V'} \mathrm{surplus}^l(y).
    \end{align*}
    The second inequality uses $s^l = s^{l-1}$, $V' \subset V$, and that $\mathrm{leftover}^l$ is $0$ on the new points and at most $\mathrm{leftover}^{l-1}$ on $W_{\mathrm{old}}$.
\end{description}
Hence, by Lemma~\ref{lemma:upper_time1_shift}, $C^l$ is reachable from $C^{l-1}$ within time~$1$.
\end{proof}

\begin{lemma}[Time of a Range Increase Ending in Phase~2]
\label{lemma:upper_redis_grow_phase2}
Let $i \in \{2, \dots, k\}$, and let $C_{i-1}$ denote the configuration of $\algtwo$ on input $I$ after step $i-1$. Assume that
\begin{align*}
    &k-(i-1) < |M_{\mathrm{maxdist}(i)}| = 2 \mathrm{maxdist}(i) + 1 \quad \text{and} \\
    &\mathrm{maxdist}(i) - \mathrm{maxdist}(i-1) = \gamma > 0.
\end{align*}
Then there exists a configuration $\hat{C}$ that is evenly distributed on $M_{\mathrm{maxdist}(i)}$ with respect to $\mathrm{hist}(\cdot, i-1)$ and is reachable from $C_{i-1}$ within time
\[
    \begin{cases}
        1 + \gamma & \text{if } k-(i-1) < |M_{\mathrm{maxdist}(i-1)}|
                     \text{ (previous request in Phase~2)}, \\
        2 + \gamma & \text{if } k-(i-1) \geq |M_{\mathrm{maxdist}(i-1)}|
                     \text{ (previous request in Phase~1)}.
    \end{cases}
\]
\end{lemma}
\begin{proof}
\label{proof:upper_redis_grow_phase2}
\leavevmode\par

\begin{description}
    \item[Previous request in Phase~2:] That is, $k-(i-1) < |M_{\mathrm{maxdist}(i-1)}|$.
    Since $\algtwo$ redistributes at every Phase~2 request by Lemma~\ref{lemma:alg2_properties}(c), the servers are evenly distributed on $M_{\mathrm{maxdist}(i-1)}$ at the end of step $i-1$. We show that moving them to the positions
    \[
        \mathit{pos}^{\star}_t \da (o - \mathrm{maxdist}(i)) + \left\lfloor \frac{(2t-1) |M_{\mathrm{maxdist}(i)}|}{2 \cdot (k-(i-1))} \right\rfloor,
        \quad t \in \{1, \dots, k-(i-1)\}
    \]
    takes at most time $1 + \gamma$, by shifting the server at
    \[
        \mathit{pos}^{i-1}_t \da (o - \mathrm{maxdist}(i-1)) + \left\lfloor \frac{(2t-1) |M_{\mathrm{maxdist}(i-1)}|}{2 \cdot (k-(i-1))} \right\rfloor
    \]
    to $\mathit{pos}^{\star}_t$. We bound this distance:
    \begin{align*}
        &\mathit{pos}^{i-1}_t - \mathit{pos}^{\star}_t \\
        &= \gamma + \left\lfloor \frac{(2t-1) \cdot |M_{\mathrm{maxdist}(i-1)}|}{2 \cdot (k-(i-1))} \right\rfloor-\left\lfloor \frac{(2t-1) \cdot |M_{\mathrm{maxdist}(i)}|}{2 \cdot (k-(i-1))} \right\rfloor \\
        &\leq \gamma - \frac{(2t-1) \cdot 2\gamma}{2 \cdot (k-(i-1))} + 1 \tag{$|M_{\mathrm{maxdist}(i)}| = |M_{\mathrm{maxdist}(i-1)}| + 2\gamma$}\\
        &\leq \gamma + 1 \tag{$t \geq 1$}
    \end{align*}
    and similarly
    \begin{align*}
        \mathit{pos}^{i-1}_t - \mathit{pos}^{\star}_t
        &\geq \gamma - \frac{(2t-1) \cdot 2\gamma}{2 \cdot (k-(i-1))} - 1 \\
        &\geq - \gamma - 1 \tag{$t \leq k-(i-1)$},
    \end{align*}
    so $\left| \mathit{pos}^{\star}_t - \mathit{pos}^{i-1}_t \right| \leq \gamma + 1$ for every server, and since the servers can move simultaneously, the total cost is at most $1 + \gamma$.

     \item[Previous request in Phase~1:] That is, $k-(i-1) \geq |M_{\mathrm{maxdist}(i-1)}|$.
    Let $\gamma_1$ be the largest number with $|M_{\mathrm{maxdist}(i-1) + \gamma_1}| \leq k - (i-1)$, and let $\beta \da \mathrm{maxdist}(i-1) + \gamma_1$; by the two assumptions, $0 \leq \gamma_1 < \gamma$. By maximality of $\gamma_1$, we have $|M_\beta| \in \{k-i, k-i+1\}$.

    We decompose the redistribution into three sequential movements:
    \begin{description}
        \item[(a) Within Phase~1:]
        Since $|M_\beta| \leq k-(i-1)$, the range increase up to $M_\beta$ stays in Phase~1. Hence Lemma~\ref{lemma:upper_redis_grow_phase1} with $\gamma_1$ in place of $\gamma$ gives an evenly distributed configuration on $M_\beta$ that is reachable from $C_{i-1}$ within time $1 + \gamma_1$.

        \item[(b) Phase transition:]
        If $|M_\beta| = k-i+1$, every point of $M_\beta$ holds exactly one free server, which is already the even distribution of Definition~\ref{def:evenly_distributed}; we set $\gamma_1' \da \gamma_1$. Otherwise, $|M_\beta| = k-i$, so every point of $M_\beta$ holds one free server and one point $p_z$, say without loss of generality $z \leq o$, holds a second one. Then we increase the range by one more step and set $\gamma_1' \da \gamma_1 + 1$: on $M_{\beta+1}$, the $2\beta+2$ free servers are evenly distributed if every point except $p_o$ holds exactly one of them, since Definition~\ref{def:evenly_distributed} with $f = 2\beta+2$ and $|S| = 2\beta+3$ gives $\lfloor (2t-1)|S|/(2f) \rfloor = t-1+\lfloor \frac12 + \frac{2t-1}{2f} \rfloor$, which equals $t-1$ for $t \leq \beta+1$ and $t$ otherwise. We reach this in time~$1$ by moving the second server of $p_z$ and all free servers left of $p_z$ one point to the left, and the remaining free servers on $p_o, \dots, p_{o+\beta}$ one point to the right.

        \item[(c) Within Phase~2:]
        Finally, we increase the range to $M_{\mathrm{maxdist}(i)}$, i.e., by the remaining $\gamma - \gamma_1'$ steps. Since the configuration is now Phase~2 evenly distributed, the same calculation as in the first case applies with $\gamma - \gamma_1'$ in place of $\gamma$, at cost $1 + \gamma - \gamma_1'$.
    \end{description}
    By the triangle inequality (Section~\ref{sec:prelim}), the total cost is at most the sum of the times of the three movements, i.e., at most
    \[
       (1 + \gamma_1') + (1 + \gamma - \gamma_1') = 2 + \gamma.
    \]
\end{description}
\end{proof}

\paragraph{Total cost.}

We extend the per-movement bounds of Section~\ref{sec:upper_known} to $\algtwo$: its initial distribution and its Phase~1 and Phase~2 redistributions get the same bounds as those of $\algone$, with the current range $\alpha'$ in place of $\alpha$, by Lemma~\ref{lemma:alg2_properties}(a) for Phase~1 and by~(c) and~(d) for Phase~2. In addition, a range increase in iteration $i$ by $\gamma_i \da \mathrm{maxdist}(i) - \mathrm{maxdist}(i-1) > 0$ costs at most $\tau \in \{1+\gamma_i, 2+\gamma_i\}$ (Lemmas~\ref{lemma:upper_redis_grow_phase1} and~\ref{lemma:upper_redis_grow_phase2}). Let $\overline{\cost}(I,\algtwo)$ be the sum of these bounds over all movements $\algtwo$ performs on $I$. Again, $\cost(I,\algtwo) \leq \overline{\cost}(I,\algtwo)$, since an iteration containing both a range increase and a redistribution costs at most the sum of both bounds (Section~\ref{sec:prelim}).

\begin{lemma}[Cost of Not Knowing $\alpha$]
\label{lemma:upper_unknown_additional}
For every $\alpha$-instance $I$, we have
\[
 \overline{\cost}(I, \algtwo) \leq \overline{\cost}(I, \algone) + 2\alpha + 1.
\]
\end{lemma}
\begin{proof}
\label{proof:upper_bound_alg_unknown_additional}
The idea is that, apart from range increases, $\algtwo$ behaves like $\algone$ on the smaller range $M_{\alpha'}$. A smaller range never makes a redistribution more expensive, and it makes $\algtwo$ wait at least as long until its next redistribution. Hence $\algtwo$ pays at most what $\algone$ pays, plus the cost of its range increases, which add up to at most $2\alpha+1$.

Here, a redistribution is a movement in one of the two redistribution branches; range increases are treated separately, also in iterations that contain both. We compare the bounds of both algorithms in four parts: the initial distribution, the range increases, and the redistributions in iterations $i \geq 2$ with $k-i \geq |M_\alpha|$ and with $k-i < |M_\alpha|$, in which $\algone$ is in Phase~1 and Phase~2, respectively. \medskip

\textbf{Initial distribution.} Its bound is $\mathrm{maxdist}(1)$ for $\algtwo$ and $\alpha \geq \mathrm{maxdist}(1)$ for $\algone$.\medskip

\textbf{Range increases.} A range increase by $\gamma_i$ costs at most $1+\gamma_i$; only the one in which $\algtwo$ switches from Phase~1 to Phase~2 may cost $2+\gamma_i$, and there is at most one such range increase, since $\algtwo$ remains in Phase~2 afterwards. Since the range grows by at least~$1$ with every range increase and never exceeds $\alpha$, there are at most $\alpha$ range increases, and their values $\gamma_i$ sum to at most $\alpha$. Hence all range increases together cost at most $2\alpha+1$. \medskip

\textbf{Redistributions in iterations $\boldsymbol{i \geq 2}$ with $\boldsymbol{k-i \geq |M_\alpha|}$.} In these iterations, both algorithms are in Phase~1, since $|M_{\alpha'}| \leq |M_\alpha| \leq k-i$. Hence every redistribution has bound~$1$, and it suffices to show that $\algtwo$ redistributes at most as often as $\algone$ in these iterations.

Let $g(i)$ be the iteration in which $\algone$ redistributes next after a redistribution in iteration~$i$. By the redistribution schedule of Section~\ref{sec:upper_known},
\[
g(i) = i + \max\left\{1, \left\lfloor \frac{k-i}{|M_\alpha|} \right\rfloor\right\}.
\]
Since $\lfloor (k-i)/|M_\alpha| \rfloor$ decreases by at most~$1$ when $i$ increases by~$1$, the function $g$ is non-decreasing.

We first show that $\algtwo$ waits at least as long as $\algone$: if $\algtwo$ redistributes in iteration $j$, where the initial distribution counts as a redistribution in iteration~$1$, then its next redistribution is in some iteration $i' \geq g(j)$. Let $j' \geq j$ be the most recent redistribution before $i'$, where, only for determining $j'$, a range increase counts as a redistribution in the iteration before it, as above, and let $\alpha'$ be the range in iteration $i'$. By the redistribution schedule with $\alpha'$ in place of $\alpha$, and since $|M_{\alpha'}| \leq |M_\alpha|$ and $g$ is non-decreasing,
\[
i' = j' + \max\{1, \lfloor (k-j')/|M_{\alpha'}| \rfloor\} \geq g(j') \geq g(j).
\]
It follows that the $t$-th redistribution of $\algtwo$ never takes place earlier than the $t$-th redistribution of $\algone$. Formally, let $a_1 < a_2 < \dots$ be the iterations $i \geq 2$ with $k-i \geq |M_\alpha|$ in which $\algtwo$ redistributes, where range increases do not count as redistributions, let $d_1 < d_2 < \dots$ be those in which $\algone$ redistributes, and set $a_0 = d_0 = 1$ for the initial distributions. We show by induction on~$t$ that whenever $a_t$ exists, $d_t$ exists as well and $d_t \leq a_t$. This implies that $\algtwo$ redistributes at most as often as $\algone$ in these iterations. For $t = 0$, we have $a_0 = d_0 = 1$. Now assume that $d_t \leq a_t$ and that $a_{t+1}$ exists. After $d_t$, $\algone$ redistributes next in iteration $g(d_t)$. Since $g$ is non-decreasing and $d_t \leq a_t$, and since $a_{t+1}$ is the next redistribution of $\algtwo$ after $a_t$, the previous paragraph gives
\[
g(d_t) \leq g(a_t) \leq a_{t+1}.
\]
Hence $k - g(d_t) \geq k - a_{t+1} \geq |M_\alpha|$, so this next redistribution of $\algone$ again lies in one of these iterations, i.e., $d_{t+1} = g(d_t) \leq a_{t+1}$. \medskip

\textbf{Redistributions in iterations $\boldsymbol{i \geq 2}$ with $\boldsymbol{k-i < |M_\alpha|}$.} In each such iteration, $\algone$ is in Phase~2 and redistributes with bound $\lfloor |M_\alpha|/(k-i+1) \rfloor + 1$. It therefore suffices to show that the bound for $\algtwo$ in the same iteration is at most this value. $\algtwo$ redistributes at most once per iteration, and if it does, there are two cases. If $\algtwo$ is still in Phase~1, which can happen since its range may be smaller, its bound is $1$, which is at most the bound for $\algone$. If $\algtwo$ is also in Phase~2, its bound is $\lfloor |M_{\alpha'}|/(k-i+1) \rfloor + 1$, which is at most the bound for $\algone$ since $|M_{\alpha'}| \leq |M_\alpha|$. For $i = k$, the same argument works with the bounds $\lfloor |M_\alpha|/2 \rfloor + 1$ and $\lfloor |M_{\alpha'}|/2 \rfloor + 1$.

Adding the bounds of all four parts proves the lemma.
\end{proof}

\paragraph{Competitive ratio.}
Combining Lemmas~\ref{lemma:upper_total_small_k} and~\ref{lemma:upper_total_big_k} with Lemma~\ref{lemma:upper_unknown_additional} and dividing by $\cost(I,\Opt) = \alpha$ (Lemma~\ref{lemma:opt_cost}) yields our final upper bounds on the competitive ratio.

\begin{theorem}[Upper Bound for $k-1 \leq 2\alpha+1$]
\label{theo:upper_bound_ratio_small_k}
For $k \geq 2$ with $k-1 \leq |M_\alpha|$, $\algtwo$ is
\[
\frac{2\alpha + 1}{\alpha} \left( H(k-1) + 2\right) \text{-competitive on $\alpha$-instances.}
\]
\end{theorem}
\begin{proof}
\label{proof:upper_bound_ratio_small_k}
Combining Lemma~\ref{lemma:upper_total_small_k} and Lemma~\ref{lemma:upper_unknown_additional}, $\algtwo$ serves all requests within total time at most $(2\alpha+1)(H(k-1)+2)$. Dividing by $\cost(I,\Opt) = \alpha$ (Lemma~\ref{lemma:opt_cost}) gives
\[
\frac{\cost(I, \algtwo)}{\cost(I, \Opt)} \leq \frac{2\alpha + 1}{\alpha} \left( H(k-1) + 2\right).
\]
\end{proof}

\begin{theorem}[Upper Bound for $k-1 > 2\alpha+1$]
\label{theo:upper_bound_ratio_big_k}
For $k-1 > |M_\alpha|=2\alpha + 1$, $\algtwo$ is
\[
\frac{\varphi_\alpha(k) + (2\alpha+1)H(2\alpha+1) + 4\alpha + 1}{\alpha} \text{-competitive on $\alpha$-instances,}
\]
where $\varphi_\alpha(k)$ is defined as in Lemma~\ref{lemma:upper_bound_phase1}.
\end{theorem}
\begin{proof}
\label{proof:upper_bound_ratio_big_k}
Combining Lemma~\ref{lemma:upper_total_big_k} and Lemma~\ref{lemma:upper_unknown_additional}, $\algtwo$ serves all requests within total time at most $\varphi_\alpha(k) + (2\alpha+1)H(2\alpha+1) + 4\alpha + 1$. Dividing by $\cost(I,\Opt) = \alpha$ (Lemma~\ref{lemma:opt_cost}) proves the theorem.
\end{proof}

\section{Randomized Bounds}\label{sec:rand_lb}

Every deterministic online algorithm is also a randomized one, so the upper bounds of Section~\ref{sec:upper_bound} also hold for randomized online algorithms. In this section, we complement them by lower bounds for randomized online algorithms.

We use Yao's principle~\cite{yao_probabilistic_1977}, \cite[Thm.~2.11]{komm_introduction_2016}: if every deterministic online algorithm has an expected cost of at least $c$ on a fixed probability distribution over instances, then for every randomized online algorithm, some instance in the support of this distribution has an expected cost of at least~$c$. Since $\cost(I,\Opt) = \alpha$ for every $\alpha$-instance $I$ (Lemma~\ref{lemma:opt_cost}), this also bounds the competitive ratio.

\paragraph{The adversary.}
Our adversary defines such a distribution: it chooses its requests at random, without observing the configuration of the algorithm. As in Section~\ref{sec:lower}, it first requests a point at distance $\alpha$ from the origin and afterwards only points of $M_\alpha$.

The following pseudocode gives the randomized adversary described above. For the first request, any algorithm incurs a cost of at least $\alpha$, as in the deterministic case. For the remaining requests, we analyze the two phases separately below.

\begin{algorithm}[H]
\caption{Randomized Adversary Strategy}\label{alg:rand_adversary}
\begin{algorithmic}[1]
\Procedure{RandomizedAdversary}{$o, k, \alpha$}
    \State $r_1 \gets p_{o+\alpha}$ ; \quad $j \gets 2$
        \Comment{distance $\alpha$ from the origin}
    \While{$j \leq k$}
        \If{$k-(j-1) \geq |M_\alpha|$} \Comment{Phase~1}
            \State $x \gets \frac{k-(j-1)}{|M_\alpha|}$
                \Comment{bound on expected surplus at a random point}
            \State choose $p \in M_\alpha$ uniformly at random
            \For{$t = 1$ \textbf{to} $\left\lceil 2x \right\rceil$}
                \If{$k-(j-1) < |M_\alpha|$} \textbf{break} \Comment{stop when Phase~2 begins}
                \EndIf
                \State $r_j \gets p$; \quad $j \gets j+1$
            \EndFor
        \Else \Comment{Phase~2}
            \State $r_j \gets$ uniformly random point in $M_\alpha$; \quad $j \gets j+1$
        \EndIf
    \EndWhile
\EndProcedure
\end{algorithmic}
\end{algorithm}

\paragraph{Phase~1.}
The adversary repeatedly chooses a point uniformly at random and requests it $\lceil 2b/|M_\alpha| \rceil$ times, where $b$ is the current number of free servers, stopping when Phase~2 begins. Since the expected number of free servers at this point is at most $b/|M_\alpha|$, by Markov's inequality, the algorithm has to move with probability at least $1/2$ in every complete round. Each complete round reduces the number of free servers by a factor of roughly $(2\alpha-1)/(2\alpha+1)$. We show that, for fixed $\alpha$, Phase~1 contains at least $(\alpha-1/2)\ln k - O(1)$ complete rounds.

\begin{lemma}[Forced Movement with Probability $\frac12$]
\label{lemma:rand_lower_prob_movement}
Assume the next request of the adversary is $r_{j+1}$ with $j \in \natp$ such that $k-j \geq |M_\alpha|=2\alpha + 1$. The adversary selects a point $p \in M_\alpha$ uniformly at random and requests it
\[
q \da \left\lceil \frac{2(k-j)}{2\alpha+1} \right\rceil
\]
times. Then $\algA$ has to move at some point within these $q$ requests with probability at least $\frac{1}{2}$.
\end{lemma}
\begin{proof}
\label{proof:rand_lower_prob_movement}
Since there are $k-j$ free servers in total,
\[
k-j = \sum_{x \in M} \mathrm{surplus}(x,j) \geq \sum_{x \in M_\alpha} \mathrm{surplus}(x,j),
\]
so $\mathbb{E}_{p \sim U(M_\alpha)}[\mathrm{surplus}(p,j)] \leq \frac{k-j}{2\alpha+1}$, where $U(M_\alpha)$ denotes the uniform distribution on $M_\alpha$.

By Markov's inequality and $q \geq \frac{2(k-j)}{2\alpha+1}$, we have
\[
\Pr\left[\mathrm{surplus}(p,j) \geq q\right]
\leq \frac{\mathbb{E}_{p \sim U(M_\alpha)}[\mathrm{surplus}(p,j)]}{q}
\leq \frac{k-j}{2\alpha+1} \cdot \frac{2\alpha+1}{2(k-j)}
= \frac{1}{2},
\]
which means that the probability that $\algA$ does not have to move within the next $q$ requests is at most $\frac{1}{2}$, i.e., it has to move with probability at least $\frac{1}{2}$.
\end{proof}

\begin{lemma}[Randomized Lower Bound for Phase~1]
\label{lemma:rand_lower_bound_phase1}
For $k > |M_\alpha| = 2\alpha+1$, the randomized adversary forces $\algA$ to pay an expected cost of at least
\[
\frac{2\alpha-1}{4} \cdot \bigl(H(k+\alpha-1) - H(3\alpha+1)\bigr) - 1
\]
to serve the requests $r_2, \dots, r_{j_1}$ with $j_1 \da k-2\alpha$.
\end{lemma}
\begin{proof}
\label{proof:rand_lower_bound_phase1}
Let
\[
\lambda \da \frac{|M_\alpha|-2}{|M_\alpha|}, \qquad b_i \da \left\lceil \lambda^i \cdot (k-1) - \sum_{t=0}^{i-1} \lambda^t \right\rceil \quad \text{for } i \geq 0.
\]
We call one execution of the Phase~1 branch of the randomized adversary a \emph{round}: the adversary picks a point $p \in M_\alpha$ uniformly at random and requests it $\lceil 2x \rceil$ times, where $x$ is the current number of free servers divided by $|M_\alpha|$. A round thus consists of several requests, and we index rounds by $i$.

We first show by induction over $i \geq 0$: if $b_i \geq |M_\alpha|$, then the first $i$ rounds are contained in the first $k-b_i$ requests and force $\algA$ to pay an expected cost of at least $\frac{i}{2}$.

\textbf{Base case ($\boldsymbol{i=0}$).} We have $b_0 = k-1$, so the claim states that the first $0$ rounds are contained in the $k-b_0 = 1$ request and force an expected cost of at least $0$, which is clearly true.

\textbf{Induction step.} Fix $i \geq 0$ with $b_{i+1} \geq |M_\alpha|$. Since $b_i \geq b_{i+1}$, as the expression inside the ceiling decreases in $i$, the induction hypothesis applies, so after the first $i$ rounds, $b \geq b_i$ requests are still to come, and $b$ is also the number of free servers. Hence the adversary is in Phase~1 and starts round $i+1$ with $q \da \lceil 2b/|M_\alpha| \rceil$ requests. We have
\begin{align*}
b - q
&\geq b \cdot \lambda - 1 \tag{$\lceil x \rceil \leq x+1$}\\
&\geq b_i \cdot \lambda - 1 \tag{$b \geq b_i$}\\
&\geq \lambda^{i+1} \cdot (k-1) - \sum_{t=0}^{i} \lambda^{t}, \tag{def $b_i$}
\end{align*}
so $b-q \geq b_{i+1} \geq |M_\alpha|$, as $b-q$ is an integer. Thus round $i+1$ is completed before Phase~2 begins, and it lies within the first $k-b_{i+1}$ requests. Since its point is chosen independently of all previous requests, Lemma~\ref{lemma:rand_lower_prob_movement} shows that $\algA$ has to move during this round with probability at least $\frac12$, which costs at least $1$. This gives an expected cost of at least $\frac{i+1}{2}$ for the first $i+1$ rounds.

We now determine how large $i$ can become while $b_i \geq |M_\alpha|$ still holds. In the first step, we use that $b_i$ is an integer and evaluate the geometric sum:
\begin{align*}
b_{i} < |M_\alpha|
&\implies \lambda^i \cdot (k-1) - \frac{1-\lambda^i}{1-\lambda} \leq |M_\alpha|-1 \tag{def.\ $b_i$}\\
&\implies \lambda^i \cdot \left(k-1+\frac{|M_\alpha|}{2}\right) \leq \frac{3}{2}|M_\alpha|-1 \tag{$\frac{1}{1-\lambda} = \frac{|M_\alpha|}{2}$}\\
&\implies \frac{k-1+|M_\alpha|/2}{\frac{3}{2}|M_\alpha|-1} \leq \left(\frac{1}{\lambda}\right)^i \\
&\implies \frac{\ln\left(\frac{k-1+|M_\alpha|/2}{\frac{3}{2}|M_\alpha|-1}\right)}{\ln\left(1+\frac{2}{|M_\alpha|-2}\right)} \leq i \tag{$\tfrac{1}{\lambda} = 1+\tfrac{2}{|M_\alpha|-2}$}\\
&\implies \frac{|M_\alpha|-2}{2} \cdot \ln\left(\frac{k-1+|M_\alpha|/2}{\frac{3}{2}|M_\alpha|-1}\right) \leq i \tag{$\ln(1+x) \leq x$}\\
&\implies \frac{|M_\alpha|-2}{2} \cdot \bigl(H(k+\alpha-1) - H(3\alpha+1)\bigr) \leq i \tag{$*$}
\end{align*}
In the second-to-last step, the logarithm is non-negative since $k > |M_\alpha|$.
For $(*)$ note that $\frac{1}{l} \leq \int_{l-1/2}^{l+1/2} \frac{1}{x}\,dx$ for all $l \geq 1$ by convexity of $x \mapsto \frac{1}{x}$, hence $H(m)-H(n) \leq \ln\frac{m+1/2}{n+1/2}$ for all integers $m > n \geq 1$. We apply this with $m = k+\alpha-1$ and $n = 3\alpha$, for which the right-hand side equals the logarithm in the previous line, and use $H(3\alpha+1) \geq H(3\alpha)$.

Therefore,
\[
i < \frac{|M_\alpha|-2}{2} \cdot \bigl(H(k+\alpha-1) - H(3\alpha+1)\bigr) \implies b_{i} \geq |M_\alpha|.
\]
Since $i$ is an integer, the largest value of $i$ for which the induction statement applies is at least
\[
\frac{|M_\alpha|-2}{2} \cdot \bigl(H(k+\alpha-1) - H(3\alpha+1)\bigr) - 1.
\]
This means that in expectation the adversary forces $\algA$ to pay a cost of at least
\[
\frac{|M_\alpha|-2}{4} \cdot \bigl(H(k+\alpha-1) - H(3\alpha+1)\bigr) - 1
= \frac{2\alpha-1}{4} \cdot \bigl(H(k+\alpha-1) - H(3\alpha+1)\bigr) - 1.
\]
Finally, $b_i \geq |M_\alpha|$ implies $k-b_i \leq k-2\alpha-1 < j_1$, so all these rounds lie within the requests $r_2, \dots, r_{j_1}$, which proves the lemma.
\end{proof}

\paragraph{Phase~2.}
In Phase~2, the adversary requests uniformly random points of $M_\alpha$. Since serving a request costs at least the distance to the closest free server, we first bound this distance in expectation using a counting argument as in the deterministic case.

\begin{lemma}[Expected Distance to the Closest Free Server]
\label{lemma:rand_lower_dist_phase2}
Let $S \subseteq M$ with $1 \leq |S| < |M_\alpha|$. Then
\[
\mathbb{E}_{p \sim U(M_\alpha)}\Bigl[\min_{s \in S} d(s,p)\Bigr] > \frac{2\alpha+1}{4|S|} - \frac{1}{4}.
\]
\end{lemma}
\begin{proof}
\label{proof:rand_lower_dist_phase2}
For every $t \in \natp$, each point of $S$ has at most $2t-1$ points within distance $t-1$, so at least $|M_\alpha| - |S|(2t-1)$ points of $M_\alpha$ have distance at least $t$ from $S$. Every point of $M_\alpha$ is counted once for each $t$ up to its distance from $S$, so summing over $t = 1, \dots, q$ gives, for every $q \in \nat$,
\[
\sum_{y \in M_\alpha} \min_{s \in S} d(s,y) \geq \sum_{t=1}^{q} \bigl(|M_\alpha| - |S|(2t-1)\bigr) = q|M_\alpha| - |S|q^2 = \frac{|M_\alpha|^2}{4|S|} - |S|\Bigl(q - \frac{|M_\alpha|}{2|S|}\Bigr)^2.
\]
Choosing $q$ as an integer closest to $\frac{|M_\alpha|}{2|S|}$, so that $\bigl|q - \frac{|M_\alpha|}{2|S|}\bigr| \leq \frac12$, gives
\[
\sum_{y \in M_\alpha} \min_{s \in S} d(s,y) \geq \frac{|M_\alpha|^2 - |S|^2}{4|S|}.
\]
Dividing by $|M_\alpha| = 2\alpha+1$ and using $|S| < |M_\alpha|$, we get
\[
\mathbb{E}_{p \sim U(M_\alpha)}\Bigl[\min_{s \in S} d(s,p)\Bigr]
\geq \frac{2\alpha+1}{4|S|} - \frac{|S|}{4(2\alpha+1)}
> \frac{2\alpha+1}{4|S|} - \frac{1}{4}.
\]
\end{proof}

\begin{lemma}[Randomized Lower Bound for Phase~2]
\label{lemma:rand_lower_bound_phase2}
Let $2 \leq j \leq k$ with $k-j+1 < |M_\alpha|$. Then the expected cost of $\algA$ to serve the requests $r_j, \dots, r_k$ is at least
\[
\frac{2\alpha + 1}{4} H(k-j+1) - \frac{k-j+1}{4}.
\]
\end{lemma}
\begin{proof}
\label{proof:rand_lower_bound_phase2}
Consider a request $r_l$ with $l \in \{j, \dots, k\}$. Before this request is served, $l-1$ servers are fixed and $k-l+1$ free servers remain. Let
\[
S \da \{ x \in M : \mathrm{surplus}(x,l-1) > 0 \}.
\]
Then $1 \leq |S| \leq k-l+1 < |M_\alpha|$, since at least one free server exists for the next request. Since the adversary chooses $r_l$ uniformly at random from $M_\alpha$, Lemma~\ref{lemma:rand_lower_dist_phase2} gives an expected distance to the closest free server, and hence an expected cost, of at least
\begin{align*}
    \frac{2\alpha+1}{4|S|} - \frac{1}{4}
    &\geq \frac{2\alpha+1}{4(k-l+1)} - \frac{1}{4}. \tag{$|S| \leq k-l+1$}
\end{align*}
Summing over all requests, the expected total cost in Phase~2 is at least
\begin{align*}
\sum_{l=j}^{k} \left( \frac{2\alpha+1}{4(k-l+1)} - \frac{1}{4}\right)
&= \frac{2\alpha+1}{4} \sum_{h=1}^{k-j+1} \frac{1}{h} - \frac{k-j+1}{4} \tag{$h=k-l+1$} \\
&= \frac{2\alpha+1}{4} H(k-j+1) - \frac{k-j+1}{4}.
\end{align*}
\end{proof}

\paragraph{Competitive ratio.}
Combining both phases with the cost $\alpha$ of the first request yields the following two theorems.

\begin{theorem}[Randomized Lower Bound for $k \leq 2\alpha+1$]
\label{theo:rand_lower_bound_ratio_small_k}
For $k \leq |M_\alpha| = 2\alpha+1$ and every randomized online algorithm $R$, there exists an $\alpha$-instance $I$ with
\[
\frac{\mathbb{E}[\cost(I,R)]}{\cost(I,\Opt)}
\geq 1 + \frac{2\alpha+1}{4\alpha} H(k-1) - \frac{k-1}{4\alpha}
\geq \frac{2\alpha+1}{4\alpha} H(k-1).
\]
\end{theorem}
\begin{proof}
\label{proof:rand_lower_bound_ratio_small_k}
By Yao's principle and since $\cost(I,\Opt) = \alpha$ for every instance the randomized adversary produces (Lemma~\ref{lemma:opt_cost}), it suffices to bound the expected cost of every deterministic online algorithm $\algA$ against this adversary from below.

To serve the first request, $\algA$ pays at least $\alpha$. For $k = 1$, this already proves the claim; so let $k \geq 2$. Since $k-1 < |M_\alpha|$, Lemma~\ref{lemma:rand_lower_bound_phase2} applies with $j=2$, giving an expected cost of at least $\frac{2\alpha+1}{4} H(k-1) - \frac{k-1}{4}$ for $r_2$ to $r_k$. Together with $r_1$, the total expected cost is at least $\alpha + \frac{2\alpha+1}{4} H(k-1) - \frac{k-1}{4}$.

Hence, for every randomized online algorithm $R$, some $\alpha$-instance $I$ produced by the adversary has at least this expected cost, and dividing by $\cost(I,\Opt) = \alpha$ gives
\[
\frac{\mathbb{E}[\cost(I,R)]}{\cost(I,\Opt)} \geq 1 + \frac{2\alpha+1}{4\alpha} H(k-1) - \frac{k-1}{4\alpha}.
\]
Since $k \leq |M_\alpha|=2\alpha+1$ implies $\frac{k-1}{4\alpha} \leq \frac12$, this is at least $\frac{2\alpha+1}{4\alpha} H(k-1)$.
\end{proof}

\begin{theorem}[Randomized Lower Bound for $k > 2\alpha+1$]
\label{theo:rand_lower_bound_ratio_big_k}
For $k > |M_\alpha| = 2\alpha+1$ and every randomized online algorithm $R$, there exists an $\alpha$-instance $I$ with
\[
\frac{\mathbb{E}[\cost(I,R)]}{\cost(I,\Opt)} \geq \frac{2\alpha-1}{4\alpha} \bigl(H(k+\alpha-1) - H(3\alpha+1)\bigr) + \frac{2\alpha+1}{4\alpha} H(2\alpha) + \frac{\alpha-2}{2\alpha}.
\]
\end{theorem}
\begin{proof}
\label{proof:rand_lower_bound_ratio_big_k}
As in the proof of Theorem~\ref{theo:rand_lower_bound_ratio_small_k}, it suffices by Yao's principle to bound the expected cost of every deterministic online algorithm $\algA$ against the randomized adversary from below.

To serve the first request, $\algA$ pays at least $\alpha$. Since $k-1 \geq |M_\alpha|$, Lemma~\ref{lemma:rand_lower_bound_phase1} gives a Phase~1 cost of at least $\frac{2\alpha-1}{4} \cdot \bigl(H(k+\alpha-1) - H(3\alpha+1)\bigr) - 1$. At the start of Phase~2, exactly $|M_\alpha|-1$ requests remain, so Lemma~\ref{lemma:rand_lower_bound_phase2} (with $k-j+1=|M_\alpha|-1=2\alpha$) gives a Phase~2 cost of at least $\frac{2\alpha+1}{4} H(2\alpha) - \frac{\alpha}{2}$. Adding these three bounds gives a total expected cost of at least
\begin{align*}
&\alpha - 1 - \frac{\alpha}{2} + \frac{2\alpha-1}{4}\bigl(H(k+\alpha-1)-H(3\alpha+1)\bigr) + \frac{2\alpha+1}{4}H(2\alpha) \\
&= \frac{2\alpha-1}{4}\bigl(H(k+\alpha-1)-H(3\alpha+1)\bigr) + \frac{2\alpha+1}{4}H(2\alpha) + \frac{\alpha-2}{2}.
\end{align*}

Hence, for every randomized online algorithm $R$, some $\alpha$-instance $I$ produced by the adversary has at least this expected cost, and dividing by $\cost(I,\Opt) = \alpha$ gives
\[
\frac{\mathbb{E}[\cost(I,R)]}{\cost(I,\Opt)} \geq \frac{2\alpha-1}{4\alpha} \bigl(H(k+\alpha-1) - H(3\alpha+1)\bigr) + \frac{2\alpha+1}{4\alpha} H(2\alpha) + \frac{\alpha-2}{2\alpha}.
\]
\end{proof}

In particular, every randomized online algorithm has a competitive ratio of at least $\frac14 H(k-1)$. Since every deterministic online algorithm is also a randomized one, Theorems~\ref{theo:upper_bound_ratio_small_k} and~\ref{theo:upper_bound_ratio_big_k} also bound the best randomized competitive ratio.

\begin{proof}[Proof of Theorem~\ref{theo:main}]
\label{proof:main}
\leavevmode\par
\textbf{Part~(i), deterministic lower bound.} If $k \leq 2\alpha+1$, Theorem~\ref{theo:lower_bound_ratio_small_k} gives a competitive ratio of at least $\frac{2\alpha+1}{2\alpha}H(k-1) \geq H(k-1)$. If $k > 2\alpha+1$, Theorem~\ref{theo:lower_bound_ratio_big_k} gives a competitive ratio of at least
\[
\frac{\mu_\alpha}{\alpha}\bigl(H(k+2\alpha) - H(4\alpha+1)\bigr) + H(2\alpha) - \frac{1}{\alpha}.
\]
Since $\frac{\mu_\alpha}{\alpha} > 2$ and $i+2\alpha+1 \leq 2i$ for $i \geq 2\alpha+1$, we have
\begin{align*}
\frac{\mu_\alpha}{\alpha}\bigl(H(k+2\alpha) - H(4\alpha+1)\bigr)
&= \frac{\mu_\alpha}{\alpha} \sum_{i=2\alpha+1}^{k-1} \frac{1}{i+2\alpha+1} \\
&\geq \sum_{i=2\alpha+1}^{k-1} \frac{1}{i} = H(k-1) - H(2\alpha),
\end{align*}
so the competitive ratio is at least $H(k-1) - \frac{1}{\alpha}$. \medskip

\textbf{Part~(i), upper bound.} We show that $\algtwo$ is $\bigl(\frac{2\alpha+1}{\alpha}H(k-1) + 6 + \frac{2}{\alpha}\bigr)$-competitive, which is at most $3H(k-1)+8$ since $\alpha \geq 1$. If $k-1 \leq 2\alpha+1$, Theorem~\ref{theo:upper_bound_ratio_small_k} gives a competitive ratio of at most
\[
\frac{2\alpha+1}{\alpha}\bigl(H(k-1)+2\bigr) = \frac{2\alpha+1}{\alpha}H(k-1) + 4 + \frac{2}{\alpha}.
\]
Now let $k-1 > 2\alpha+1$, so that $H(2\alpha+1) \leq H(k-1)$. If $k \leq 4\alpha+2$, then $\varphi_\alpha(k) = k-2\alpha-2 \leq 2\alpha$, and Theorem~\ref{theo:upper_bound_ratio_big_k} gives a competitive ratio of at most
\[
\frac{2\alpha + (2\alpha+1)H(2\alpha+1) + 4\alpha+1}{\alpha} \leq \frac{2\alpha+1}{\alpha}H(k-1) + 6 + \frac{1}{\alpha}.
\]
If $k \geq 4\alpha+3$, then $\ln\frac{m}{n} = \int_n^m \frac{1}{x}\,dx \leq H(m-1) - H(n-1)$ for integers $m \geq n \geq 1$, applied with $m = k-2\alpha-1$ and $n = 2\alpha+2$, gives
\[
\varphi_\alpha(k) \leq \mu_\alpha \bigl(H(k-2\alpha-2) - H(2\alpha+1)\bigr) + 2\alpha+1.
\]
Since $H(k-2\alpha-2) - H(2\alpha+1) \geq 0$ and $\mu_\alpha < 2\alpha+1$, Theorem~\ref{theo:upper_bound_ratio_big_k} gives a competitive ratio of at most
\begin{align*}
&\frac{2\alpha+1}{\alpha}\bigl(H(k-2\alpha-2) - H(2\alpha+1)\bigr) + \frac{2\alpha+1}{\alpha}H(2\alpha+1) + \frac{6\alpha+2}{\alpha} \\
&= \frac{2\alpha+1}{\alpha}H(k-2\alpha-2) + 6 + \frac{2}{\alpha} \leq \frac{2\alpha+1}{\alpha}H(k-1) + 6 + \frac{2}{\alpha}.
\end{align*} 

\textbf{Part~(i), randomized bounds.} If $k \leq 2\alpha+1$, then $\frac{k-1}{4\alpha} \leq \frac12$, so Theorem~\ref{theo:rand_lower_bound_ratio_small_k} gives a competitive ratio of at least
\[
1 + \frac{2\alpha+1}{4\alpha} H(k-1) - \frac{k-1}{4\alpha} \geq \frac12 H(k-1) + \frac12 \geq \frac14 H(k-1).
\]
If $k > 2\alpha+1$, then $H(k+\alpha-1) \geq H(3\alpha+1)$, so by $\frac{2\alpha-1}{4\alpha} \geq \frac14$ and $H(k+\alpha-1) \geq H(k-1)$, Theorem~\ref{theo:rand_lower_bound_ratio_big_k} gives a competitive ratio of at least
\[
\frac14 \bigl(H(k+\alpha-1) - H(3\alpha+1)\bigr) + \frac{2\alpha+1}{4\alpha} H(2\alpha) + \frac{\alpha-2}{2\alpha}
\geq \frac14 H(k-1) + \Delta,
\]
where
\[
\Delta \da \frac12 H(2\alpha) - \frac14 H(3\alpha+1) + \frac{H(2\alpha)}{4\alpha} + \frac12 - \frac1\alpha.
\]
For $\alpha \geq 2$, we have $\frac12 - \frac1\alpha \geq 0$ and
\begin{align*}
H(3\alpha+1) - H(2\alpha) &\leq \ln\frac{3\alpha+1}{2\alpha} \leq \ln 2 < 1 \leq H(2\alpha), \\
\text{so}\quad \Delta &\geq \frac14\bigl(H(2\alpha) - (H(3\alpha+1) - H(2\alpha))\bigr) \geq 0.
\end{align*}
The upper bound of $3H(k-1)+8$ also holds for the best randomized competitive ratio, since every deterministic online algorithm is also a randomized one. \medskip

\textbf{Part~(ii).} Let $\alpha$ be fixed, let $k \geq 4\alpha+3$, and let $O(1)$ denote terms that depend only on $\alpha$. We use $\ln(n+1) \leq H(n) \leq \ln n + 1$ for $n \geq 1$. For the lower bound, $H(k+2\alpha) \geq \ln k$ and $H(4\alpha+1) \leq \ln(4\alpha+1) + 1$, so Theorem~\ref{theo:lower_bound_ratio_big_k} gives a competitive ratio of at least
\[
\frac{\mu_\alpha}{\alpha}\bigl(\ln k - \ln(4\alpha+1) - 1\bigr) + H(2\alpha) - \frac1\alpha = \frac{\mu_\alpha}{\alpha}\ln k - O(1).
\]
For the upper bound, $\varphi_\alpha(k) \leq \mu_\alpha \ln k + 2\alpha + 1$ since $\frac{k-2\alpha-1}{2\alpha+2} \leq k$, so by Theorem~\ref{theo:upper_bound_ratio_big_k}, the competitive ratio of $\algtwo$ is at most
\[
\frac{\mu_\alpha \ln k + (2\alpha+1)H(2\alpha+1) + 6\alpha + 2}{\alpha} = \frac{\mu_\alpha}{\alpha}\ln k + O(1).
\]
Finally, $2 < \frac{\mu_\alpha}{\alpha} < 2+\frac1\alpha$ follows from $2\alpha < \mu_\alpha < 2\alpha+1$ (Section~\ref{sec:prelim}).
\end{proof}

\section{Discussion and Open Problems}\label{sec:cycle}
By Theorem~\ref{theo:main}, randomization improves the competitive ratio by at most a constant factor. The factor-$2$ gap for $k \leq 2\alpha+1$ persists as $\alpha \to \infty$: the bounds of Theorems~\ref{theo:lower_bound_ratio_small_k} and~\ref{theo:upper_bound_ratio_small_k} tend to $1+H(k-1)$ and $2H(k-1)+4$, respectively.

Our upper bound loses this factor in the redistribution step. With $f$ free servers, serving a request costs at most about $|M_\alpha|/(2f)$, which matches our lower bound, but restoring the even distribution can cost up to about $|M_\alpha|/f$, since the distribution is pinned to both ends of $M_\alpha$. On a cycle, there are no ends, and the new distribution can be rotated so that every server moves at most half a gap. To examine the role of these two ends, we consider the continuous cycle of circumference $2\alpha$. We restrict attention to instances with a request opposite to the origin. This allows us to compare the bounds on the line with those on a metric without boundary points. Intuitively, as $\alpha \to \infty$, rounding to integer distances becomes negligible relative to $\cost(I,\Opt) = \alpha$, so the line behaves like a continuous segment of length~$2\alpha$, and the continuous cycle of circumference~$2\alpha$ differs from it only by having no ends.

\begin{theorem}[Tight Bound on the Continuous Cycle]
\label{theo:continuous_cycle}
Let $\alpha > 0$ be real. On the continuous cycle of circumference $2\alpha$, restricted to instances containing a request opposite to the origin, the best competitive ratio of deterministic online algorithms is exactly $1+H(k-1)$.
\end{theorem}
\begin{proof}
\label{proof:continuous_cycle}
Fix a real $\alpha > 0$ and a continuous cycle of circumference $2\alpha$ with origin $p_o$. Every point of the cycle has distance at most $\alpha$ from $p_o$, and the point opposite to the origin has distance exactly $\alpha$. As in the proof of Lemma~\ref{lemma:opt_cost}, the optimal offline cost is therefore $\alpha$ for every instance we consider.

\textbf{Lower bound.}
The adversary first requests the point opposite to the origin. To serve this request, a server has to travel from $p_o$ to this point, so every online algorithm pays at least $\alpha$.

Now consider a request $r_l$ with $l \in \{2, \dots, k\}$. Before this request is served, $l-1$ servers are fixed and $x \da k-l+1$ servers are free. These $x$ free servers divide the cycle into at most $x$ arcs whose lengths sum up to $2\alpha$, so at least one of these arcs has length at least $\frac{2\alpha}{x}$. Its midpoint has distance at least $\frac{\alpha}{x}$ from every free server, and the adversary requests it. Summing over all requests, the cost is at least
\begin{align*}
\alpha + \sum_{l=2}^{k} \frac{\alpha}{k-l+1}
&= \alpha + \alpha \sum_{x=1}^{k-1} \frac{1}{x} \tag{$x = k-l+1$}\\
&= \alpha \bigl(1 + H(k-1)\bigr).
\end{align*}

\textbf{Upper bound.}
We use an online algorithm that keeps the free servers evenly distributed on the whole cycle. In the first step, it serves $r_1$ and places the $k-1$ free servers evenly on the cycle. Since every point has distance at most $\alpha$ from $p_o$, the cost of this step is at most $\alpha$.

Now consider a request $r_l$ with $l \in \{2, \dots, k\}$. We maintain the invariant that before each request $r_l$ with $l \geq 2$, the $x \da k-l+1$ free servers are evenly distributed, i.e., the gaps between consecutive free servers along the cycle have length $g \da \frac{2\alpha}{x}$, where for $x=1$ the single gap is the whole cycle. After the first step this holds, and the redistribution below preserves it. If $x=1$, the last request costs at most $\alpha$ and no redistribution is needed. Otherwise, the algorithm has two simultaneous movements. \medskip

\emph{Serving the request.} Every point lies between two neighboring free servers, so its distance to the closer one is at most $\frac{g}{2} = \frac{\alpha}{x}$. The closest free server serves $r_l$ at a cost of at most $\frac{\alpha}{x}$.

\emph{Redistribution.} If $x \geq 2$, the remaining $x-1$ free servers have to be distributed evenly again, i.e., with distance $g' \da \frac{2\alpha}{x-1}$. Measured from the serving server along the cycle (i.e., renaming the positions such that the serving server was at position $0$), they are located at
\[
a_j \da j \cdot g \quad \text{for } j \in \{1, \dots, x-1\},
\]
and we move server $j$ to
\[
b_j \da \Bigl(j - \frac12\Bigr) \cdot g' \quad \text{for } j \in \{1, \dots, x-1\}.
\]
These positions are again evenly spaced, since consecutive $b_j$ differ by $g'$ and the gap from $b_{x-1}$ across position $0$ to $b_1$ is $2\alpha - (x-2)g' = g'$. The distance server $j$ has to move is
\begin{align*}
|b_j - a_j|
&= \left| \Bigl(j - \frac12\Bigr) \cdot g' - j \cdot g \right| \\
&= \left|\Bigl(j - \frac12\Bigr)(g' - g) - \frac{g}{2} \right|\\
&= \left|\Bigl(j - \frac12\Bigr) \cdot \frac{g}{x-1} - \frac{g}{2} \right| \tag{$g' - g = \frac{2\alpha}{x(x-1)} = \frac{g}{x-1}$}\\
&= \frac{|2j - x|}{2(x-1)} \cdot g \\
&\leq \frac{x-2}{2(x-1)} \cdot g \tag{$j \in \{1, \dots, x-1\}$}\\
&< \frac{g}{2} \tag{$x-2 < x-1$}\\
&= \frac{\alpha}{x}.
\end{align*}
Since the two movements involve distinct servers, they can be performed simultaneously, so the request $r_l$ costs at most $\frac{\alpha}{x}$. Summing over all requests, the total cost is at most
\begin{align*}
\alpha + \sum_{l=2}^{k} \frac{\alpha}{k-l+1}
= \alpha \bigl(1 + H(k-1)\bigr). \tag{as above}
\end{align*}

Dividing both bounds by the optimal offline cost $\alpha$ proves the theorem.
\end{proof}

Closing the factor-$2$ gap on the line remains open, as does the question whether randomization improves the constant factor in front of $H(k-1)$. A natural next step is to drop the fixed-server assumption while keeping requests cumulative. With exactly $k$ requests, our upper bounds and our Phase~1 lower bounds still hold in this setting, but our Phase~2 arguments do not transfer. More importantly, since every request fixes a server in our model, dropping this assumption is necessary to allow more than $k$ requests, and thus to move towards the general time-optimal $k$-server problem with arbitrarily long request sequences. Further directions are an arbitrary initial configuration and other metric spaces, such as trees.

\section*{Acknowledgments}
This paper is based on my bachelor's thesis. I thank Juraj Hromkovič for proposing its topic and for co-supervising it, and Dennis Komm for supervising it. I thank both of them for their valuable guidance and feedback throughout the thesis and for their support in developing it into the present paper. I also thank Peter Rossmanith for his support with the randomized lower bounds, and Hans-Joachim Böckenhauer for his helpful feedback on the presentation and the formal details of this paper.

\section*{AI Use Statement}
All results and proofs in this paper were developed and written by me. I used Claude (Claude Fable 5 and Claude Opus 5.5, Anthropic) to check proofs and calculations for errors and gaps, to suggest shorter formulations of some arguments, which I adopted in a few places after verifying them, to improve language, structure, and consistency of notation, and to help with the \LaTeX{} source. I verified all suggestions and take full responsibility for the content of the paper.

\bibliographystyle{splncs04}
\bibliography{references}

\end{document}